\documentclass[fleqn,usenatbib]{mnras}

\usepackage{newtxtext,newtxmath}

\usepackage[T1]{fontenc}

\DeclareRobustCommand{\VAN}[3]{#2}
\let\VANthebibliography\thebibliography
\def\thebibliography{\DeclareRobustCommand{\VAN}[3]{##3}\VANthebibliography}

\usepackage{xcolor}

\usepackage{graphicx}	
\usepackage{amsmath}	
\usepackage[normalem]{ulem}

\title[The MeerKAT DEEP2 field at S-band]{A first glimpse at the MeerKAT DEEP2 field at S-band}

\author[S. Ranchod et al.]{S. Ranchod,$^{1}$\thanks{E-mail: sranchod@mpifr-bonn.mpg.de}
J.~D. Wagenveld,$^{1}$
H.-R. Kl\"{o}ckner,$^{1}$
O. Wucknitz,$^{1}$
R. P. Deane,$^{2,3}$
S.~S. Sridhar,$^{4,1}$
E. Barr,$^{1}$ \newauthor S. Buchner,$^{5}$
F. Camilo,$^{5}$
A.~Damas-Segovia,$^{1}$
C. Kasemann,$^{1}$
M. Kramer,$^{1}$
L. S. Legodi,$^{5}$
S. A. Mao,$^{1}$ \newauthor
K. Menten,$^{1}$
I. Rammala,$^{1}$
M.~R.~Rugel,$^{1,6,7}$
and G. Wieching$^{1}$
\\
$^{1}$Max-Planck Institut fur Radioastronomie, Auf dem H\"{u}gel 69, 53121 Bonn, Germany\\
$^{2}$Wits Centre for Astrophysics, School of Physics, University of the Witwatersrand, 1 Jan Smuts Avenue, Johannesburg, 2000, South Africa\\
$^{3}$Department of Physics, University of Pretoria, Private Bag X20, Pretoria 0028, South Africa\\
$^{4}$SKA Observatory, Jodrell Bank, Lower Withington, Macclesfield, SK11 9FT, United Kingdom\\
$^{5}$South African Radio Astronomy Observatory, River Park, Gloucester Road Liesbeek House, Mowbray 7700, South Africa\\
$^{6}$Center for Astrophysics | Harvard \& Smithsonian, 60 Garden Street, Cambridge, MA 02138, USA\\
$^{7}$National Radio Astronomy Observatory, PO Box O, 1003 Lopezville Road, Socorro, NM 87801, USA 
}

\date{Accepted 2024 December 11. Received 2024 December 11; in original form 2024 November 04}

\pubyear{\the\year{}}

\begin{document}
\label{firstpage}
\pagerange{\pageref{firstpage}--\pageref{lastpage}}
\maketitle

\begin{abstract}
We present the first widefield extragalactic continuum catalogue with the MeerKAT S-band (2.5 GHz), of the radio-selected DEEP2 field. The combined image over the S1 (1.96 -- 2.84 GHz) and S4 (2.62 -- 3.50 GHz) sub-bands has an angular resolution of 6.8”$\times$3.6” (4.0”$\times$2.4”) at a robust weighting of $R = 0.3$  ($R=-0.5$) and a sensitivity of 4.7 (7.5) \textmu Jy~beam$^{-1}$ with an on-source integration time of 70 minutes and a minimum of 52 of the 64 antennas, for respective observations. We present the differential source counts for this field, as well as a morphological comparison of resolved sources between S-band and archival MeerKAT L-band images. We find consistent source counts with the literature and provide spectral indices fitted over a combined frequency range of 1.8 GHz. These observations provide an important first demonstration of the capabilities of MeerKAT S-band imaging with relatively short integration times, as well as a comparison with existing S-band surveys, highlighting the rich scientific potential with future MeerKAT S-band surveys.
\end{abstract}




\section{Introduction}

The DEEP2 field, defined by \citet{Mauch_2020}, is an extragalactic legacy field, selected for its local sparsity of bright radio sources. This field is the only legacy field selected based on radio properties alone and the reduced contribution from bright radio sources maximises the attainable sensitivity. This, along with its favourable circumpolar position, allows it to serve as an optimal pilot field for the evaluation of new radio observatories and instrumentation in the southern hemisphere e.g. MeerKAT, SKA, with the caveat of poor access for ground-based optical and infrared observatories.

\citet{Mauch_2020} carried out 128-hour observations on this field at 1.28 GHz with MeerKAT, which resulted in a confusion-limited observation with $\sigma = 0.55$~\textmu Jy~beam$^{-1}$, and the deepest radio continuum image at L-band frequencies. These results have contributed to the computation of source counts in the DEEP2 field over a large range of flux densities \citep{Matthews_2021} and to the determination of the probability distribution of source confusion $P(D)$ down to 0.25~\textmu Jy, allowing for new constraints on the star-formation rate of the Universe \citep{Matthews_2021b}. 

The $\sim$3~GHz regime provides a window through which to directly study feedback in AGN and to observe star-formation in galaxies without dust obscuration \citep[e.g.][]{Condon_1992,Smolcic_2017,Leslie_2020}, both of which are important to constrain galaxy formation and evolution models. At 3~GHz, the spectral energy distributions of nearby normal galaxies are dominated by synchrotron emission, with only $\sim25$\% attributed to free-free (thermal) emission \citep{Condon_1992}. While for the faint source population, \citet{Algera_2020} have shown that at $S_\mathrm{3 GHz} < 50$~\textmu Jy, the fraction of sources dominated by thermal emission increases to $\sim 90$\%. Through the radio-infrared relation, radio luminosity can be used as a direct tracer for star-formation rate \citep[e.g.][]{Murphy_2009,Delhaize_2017,Heesen2019}. The high angular resolution achieved with radio interferometers at S-band (1.75 -- 3.50 GHz for MeerKAT) allows for deep observations (<1~\textmu Jy), without being limited by confusion, to study the low-flux density source population \citep[e.g.][]{Vandervlugt_2021}, and to understand the evolution of the thermal emission fraction with redshift.

We present pilot observations of the DEEP2 field, producing the first S-band extragalactic catalogue down to 16 \textmu Jy with the new S-band system on MeerKAT. This catalogue provides an S-band measurement for the deepest L-band radio image, allowing for the computation of multi-band spectral indices. This is particularly useful for the classification of sources and provides insight into their dominant emission mechanisms. Furthermore, these pilot observations provide an early science demonstrator of the performance of the MeerKAT S-band receiver in the context of extragalactic imaging, to inform future observations at S-band frequencies. 

In this paper, we present a shallow pilot catalogue of the DEEP2 field in S-band, as well as the differential source counts. This paper is organised as follows: In Section~\ref{sec:obsfull} we describe the DEEP2 observations and our calibration and imaging strategies. In Sect.~\ref{sec:deep2} we present the S-band DEEP2 image and catalogue, and in Section~\ref{sec:sourcecounts} we present the Euclidean normalised differential source counts for the DEEP2 field at 2.5~GHz. We summarise and conclude in Sect.~\ref{sec:concl}. Throughout this paper, we define the spectral index $\alpha$ as $S_\nu \propto \nu^\alpha$, where $S_\nu$ is the flux density at frequency $\nu$.

\section{Observations and data reduction}\label{sec:obsfull}

The DEEP2 field, centred on RA = 04:13:26.4, Dec $=-80.00.00.0$ (J2000), is a radio-selected legacy field in the Southern sky. It was defined by \citep{Mauch_2020} as a commissioning deep field for the MeerKAT telescope, to measure and optimise the performance of the array. The field was selected due to its low demerit score, a parameter which describes the expected gain variation due to pointing and receiver gain error for the known bright sources in the field, to quantify the maximum attainable sensitivity with the MeerKAT L-band. The low demerit score of the DEEP2 field of $D = 1.4$~mJy is three orders of magnitude smaller than that for optically selected southern deep fields e.g. COSMOS, XMM-LSS, CDFS. These properties make this a suitable choice for an early S-band catalogue, which will be used for verification ahead of future, deeper observations.

In this section, we detail the observation setup as well as the calibration and imaging methods used in this work. From these observations, we produce the first multi-band extragalactic catalogue using the MeerKAT S-band system. A detailed introduction to the new S-band receivers, including the system equivalent flux density and the sub-band division can be found in Appendix~\ref{sec:sband}.

\subsection{Observations}\label{sec:obs}
The DEEP2 field was observed with the MeerKAT interferometer \citep{Jonas_2016} as part of the S-band science verification (Project IDs SSV-20230209-0012 and SSV-20230215-0017). We observed a single pointing of the DEEP2 field in the S1 ($1968.75-2843.75$~MHz) and S4 ($2625.00-3500.00$~MHz) sub-bands with $\sim$70 minutes on source integration time each. The flux and bandpass calibrator J0408$-$6545 (PKS~0408$-$65) was observed for 8 minutes at the end of the observation, and the phase calibrator J0252$-$7104 (PKS 0252$-$71) was observed every 10 minutes. In addition, 3C138 was observed as a polarisation calibrator for 5 minutes at the start of the observation and 8 minutes at the end. The integration time per visibility was 2 seconds. Further details of the observations are summarised in Table~\ref{tab:bands}. With this choice of sub-bands, we have a maximum total bandwidth of 1531.25 MHz, with an overlapping bandwidth of 218.75 MHz.

\begin{table*}
    \centering
    \caption{Summary of DEEP2 observation details for the respective sub-bands.}
    \begin{tabular}{lll}
    \hline
    Parameters          & S1 & S4 \\
    \hline
    Observation ID      & 20230209-0012 & 20230215-0017 \\
    Observation date    & 10 March 2023 & 12 March 2023 \\
    Frequency range     & $1968.75-2843.75$~MHz & $2625.00-3500.00$~MHz \\
    Time on source (hours) & 1.17 & 1.17\\
    Number of antennas  & 52 & 55 \\
    \hline
    \end{tabular}
    \label{tab:bands}
\end{table*}

\subsection{Calibration and imaging}
\label{sec:calibration}

The processing of these observations followed the standard radio interferometric calibration procedure, in which we corrected delay, bandpass and gain variations and flagged bad visibilities in an iterative process. For calibration and imaging we used various tasks from {\sc CASA} \citep{McMullin_2007} and {\sc WSClean} \citep{Offringa_2014} respectively, as detailed below. We followed the same procedure for both observations to allow us to validate the calibration strategy and combine both sub-bands for imaging. 

Bad visibilities were flagged using {\tt flagdata} and by using an averaged waterfall spectrum of all baselines to remove frequency-dependent RFI\footnote{ \url{https://github.com/hrkloeck/DASKMSWERKZEUGKASTEN}}, after which each dataset was divided into 16 uniform spectral windows {(SPWs)}, with an equal number of channels, for further processing. This allowed for sufficient bandwidth per SPW to solve for the per-SPW selfcal solutions. Additionally, as PKS~0408$-$65 has a non-flat spectral index, it is necessary to divide the band into multiple uniform SPWs to bootstrap the flux scale calibration along the frequency axis. The primary calibrator PKS~0408$-$65 was used for delay, bandpass, and gain calibration, assuming a point source model with a flux density of 6.986~Jy at a reference frequency of 2.7~GHz\footnote{The model was set as specified in the \href{https://skaafrica.atlassian.net/wiki/spaces/ESDKB/pages/1481408634/Flux+and+bandpass+calibration\#Using-CASA-setjy-for-non-standard-flux-models}{MeerKAT Knowledge Base}, using measurements from the NASA/IPAC Extragalactic Database (NED).}. The spectral index terms were set as {\tt spix} = [$-$1.2897, $-$0.2353, 0.0861] in {\tt setjy}. {Baselines with a projected length of $< 100$~m were excluded from the calibration process}, with reference antenna m000. For completeness, we calibrated the cross-hand polarisation measurements, using a model for 3C138 with a {\tt polindex} of 0.078 and a polarisation angle of $-9.60$~deg to determine the leakage term ({\tt polcal}), cross-hand delay ({\tt gaincal}) and phase/position angle ({\tt polcal}). The polarisation data products will be presented in a future work. In a final step, all the corrections were applied to the phase calibrator source (J0252$-$7104) and the target source, after which the visibilities were further flagged using {\tt flagdata}. The final flagged percentages of the phase calibrator and target source datasets are approximately 30\% (S1) and 9\% (S4). As also seen in Fig.~\ref{fig:sefd}, this indicates that the S1 sub-band is more affected by RFI relative to S4.

\begin{figure}
    \centering
    \includegraphics[width=0.95\columnwidth]{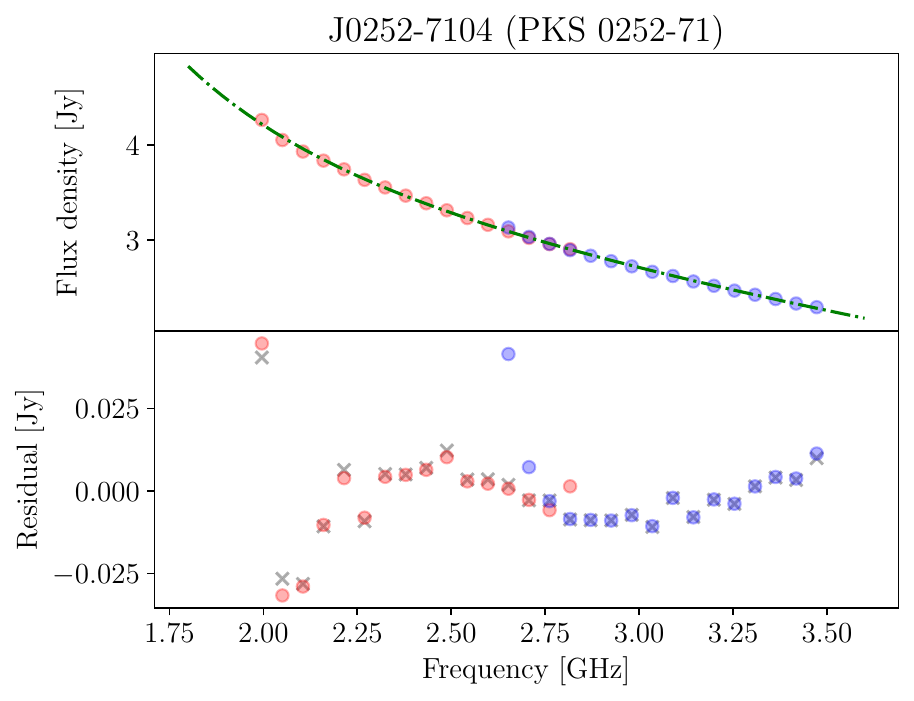}
    \caption{{\it Top:} {Flux} density of J0252$-$7104 (PKS 0252$-$71), the phase calibrator used for the observations, as a function of frequency. {\it Bottom:} The residual flux density, after applying the flux density model (green), as a function of frequency. Each marker shows the averaged measurement per SPW of both subbands (S1 in red and S4 in blue). The crosses indicates the resulting flux densities of the SPW images of the full concatenated dataset. The mean and standard deviation of the residuals are $-0.0013$ and 0.0123 Jy, respectively.}
    \label{fig:phase-cal}
\end{figure}

Initial imaging was carried out using {\sc WSClean} with a Briggs \citep{Briggs_1995} robust weighting of $R = -0.5$, following the CASA convention. Figure~\ref{fig:phase-cal} shows the flux density of the phase-calibrator source J0252$-$7104 for all SPWs of both sub-band observations. Overall, the SPWs show a smooth decrease in flux density. To investigate the differences between the sub-band observations and how the SPWs are related, a spectral model for J0252$-$7104 was determined using the same recipe as for the primary calibrator and extracted from the flux densities. The residual measurements show a standard deviation at a percentage level, while some SPWs from S1 and S4 show some slight offsets at the overlapping frequencies. These SPWs were not used and the final data set for imaging the target source was generated by combining SPWs 0 to 13 (1.97 -- 2.73 GHz) of sub-band S1 and SPWs 2 to 15 (2.73 -- 3.50 GHz) of sub-band S4.

After primary calibration and concatenating the two sub-bands, we performed phase and amplitude self-calibration using {\sc CASA} and {\sc WSClean}\footnote{\url{https://github.com/hrkloeck/2GC}}. We generated a cleaning mask\footnote{\url{https://github.com/JonahDW/Image-processing}} using the Python Blob Detector and Source Finder \citep[{\sc PyBDSF};][for more information see Section~\ref{sec:sourcefinding}]{Mohan_2015}. The resulting images ($R = -0.5$) over all 27~SPWs each have an rms (i.e. image sensitivity) of $\sigma \approx 15$~\textmu Jy~beam$^{-1}$. Through inspection of individual SPW images, we found certain SPWs had increased rms due to a higher SEFD (Fig.~\ref{fig:sefd}) or flagged percentage. We therefore selected a subset of SPWs (1--19) for the final multi-frequency synthesis (MFS) images. With this, we achieved an rms of $\sigma = 7.5$~\textmu Jy~beam$^{-1}$ for the same weighting. The final images have a frequency range of 2.02 to 3.06~GHz, covering $\sim1$~GHz bandwidth.

For further analysis we created images of two different robust weightings. We selected a weighting of $R = -0.5$ for a higher angular resolution image, a typical weighting selected for MeerKAT L-band observations. We also selected a weighting that results in higher sensitivity ($R = 0.3$) and comparable angular resolution to the \citet{Mauch_2020} L-band image of the DEEP2 field, i.e. $\theta = 7.6"\times7.6"$. The image properties for both images are summarised in Table~\ref{tab:imaging}, where the angular resolution is described by the full width at half-maximum (FWHM) of the point spread function (PSF). We present a full analysis of optimal robust parameters at S-band frequencies in Appendix~\ref{sec:robust}.

\subsection{Primary beam correction}
\label{sec:pb_correction}

Because of the primary beam response of the MeerKAT antennas, flux densities become more attenuated further from the pointing centre. To correct for the primary beam attenuation, an accurate model or measurement of the primary beam response is needed. Analytical models of the MeerKAT primary beam at S-band are available\footnote{\url{https://github.com/ska-sa/katbeam}}, however the most accurate representation of the primary beam is obtained through holographic measurements, which are also available for MeerKAT at S-band \citep{deVilliers_2023}. We therefore elected to use holographic measurements for our primary beam models. These should not differ much from the available analytical models, given that the analytical models are informed by the holographic measurements \citep{deVilliers_2022}. Because of the large bandwidths of the sub-bands (875~MHz for both S1 and S4), the frequency evolution of the primary beam must be taken into account when modeling the primary beam. The FWHM of the primary beam evolves with inverse frequency, becoming smaller at higher frequencies. As seen in Fig.~1 of \citet{deVilliers_2023}, this relation holds throughout the S-band, except for some observed peaks in the FWHM at the higher end of the band. These peaks are an expected feature of the beam elongation at the high frequency end of the MeerKAT bands. This is more pronounced at S-band, with respect to L- and UHF-band, due to its independent orthomode transducer and feed design \citep{Kramer_2016}. This primary beam elongation effect appears at the same frequencies as the peaks in the SEFD seen in Fig.~\ref{fig:sefd}. 

\begin{figure*}
    \centering
    \includegraphics[width=0.85\textwidth]{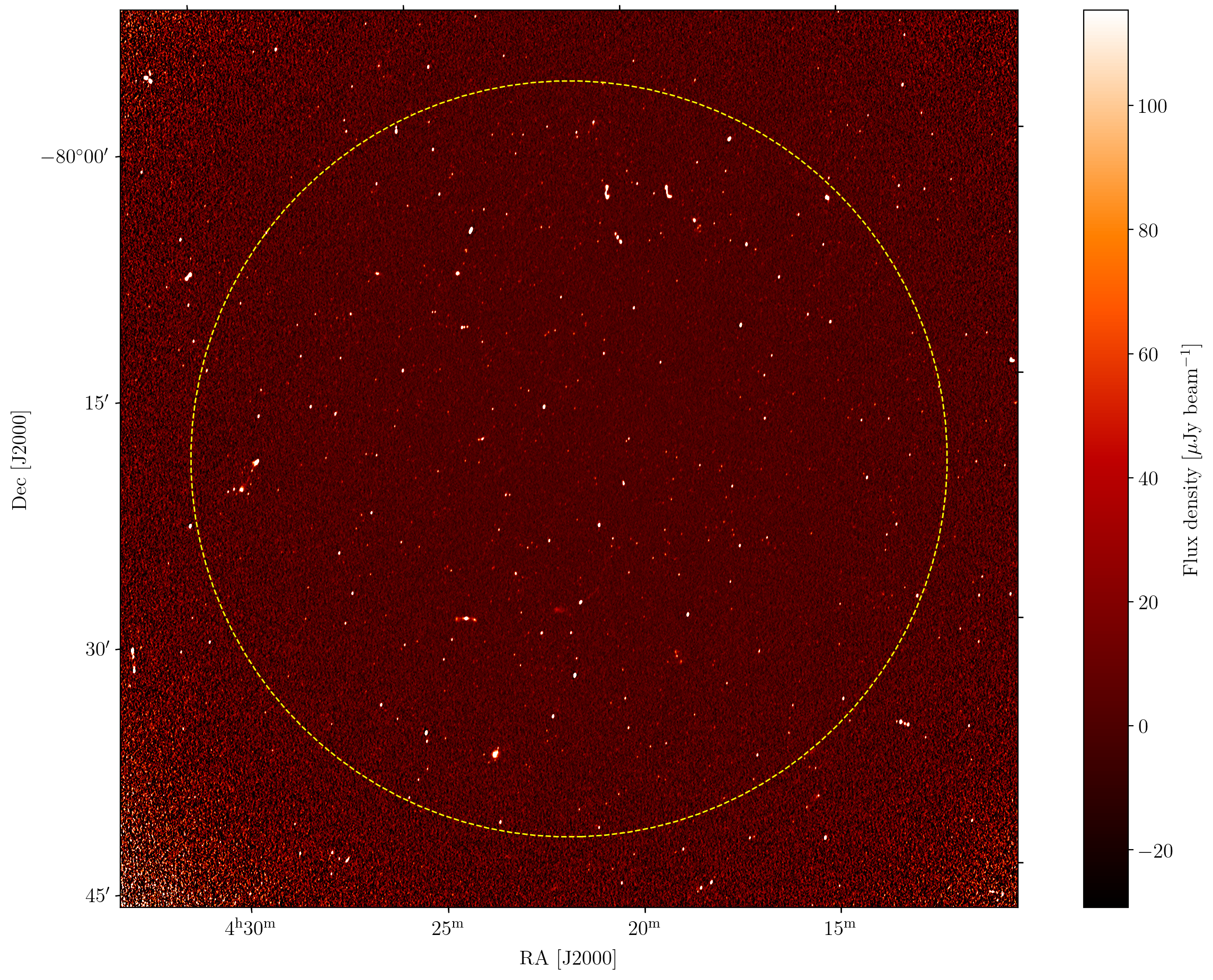}
    \caption{The central 54'$\times$54' of the combined primary beam-corrected S1- and S4-subband image of the DEEP2 field, with a Briggs robust weighting of $R=0.3$. {The FWHM of the primary beam is indicated by the yellow dashed circle, as specified in Section~\ref{sec:pb_correction}.}}
    \label{fig:deep2final}
\end{figure*}

To produce an accurate primary beam model, we also have to take into account the way in which the MFS image is produced. During imaging, the data was split into 16 SPWs per sub-band, and an MFS image was produced from the weighted average of these SPWs. As specified in Section~\ref{sec:calibration}, some of these SPWs were excluded from the final imaging, leaving 19 SPWs with which the MFS images were created. Among the excluded SPWs are those at frequencies corresponding to the peaks in FWHM seen in Fig.~1 of \citet{deVilliers_2023}. As such, we do not expect to be affected by these in the primary beam correction. To accurately represent the primary beam of the MFS images, we created a set of primary beam models for the individual SPWs and combined them into a MFS primary beam using the same weights used to create the MFS images. While applying the primary beam model, we cut off the image at the 2\% level of the primary beam, producing a nearly circular image with a diameter of {1.54}~degrees. The central $54^{\prime}\times54^{\prime}$ of the primary beam corrected $R = 0.3$ image (where, a diameter of $54^{\prime}$ corresponds to the 30\% level of the primary beam), along with the FHWM of the primary beam, is shown in Fig.~\ref{fig:deep2final}. Because of the frequency evolution of the primary beam, the S1+S4 combined image has a smaller primary beam than if we were to just use S1. The primary beam-corrected images at $R=0.3$ and $R=-0.5$ are available in FITS format on the SARAO archive\footnote{\url{https://archive.sarao.ac.za/}} at \url{https://doi.org/10.48479/zdyz-8342}. 

For an accurate comparison with the L band image in Section~\ref{sec:resolved}, we must correct the L band image retrieved from the archive for primary beam attenuation as well. To create an accurate primary beam model, we apply the same method of primary beam correction as described above, using the sub-band frequencies and weights specified in Table~5 of \citet{Mauch_2020}.

\begin{table}
    \centering
    \caption{Summary of the resulting image properties for the respective robust weightings}
    \begin{tabular}{lll}
    \hline
    Parameters & $R = -0.5$ & $R = 0.3$ \\
    \hline
    Sensitivity [\textmu Jy beam$^{-1}$] & 7.5 & 4.7 \\
    Dynamic range & 2160:1 & 3615:1\\
    PSF (maj. $\times$ min.) & $4.0" \times 2.4"$ & $6.8"\times3.6"$\\
    PSF PA (deg) & $-1.6$ & $-11.0$ \\
    N sources & 670 & 1199 \\
    \hline
    \end{tabular}
    \label{tab:imaging}
\end{table}

\section{DEEP2 image and cataloguing}\label{sec:deep2}

\subsection{Source detection}
\label{sec:sourcefinding}

To generate continuum source catalogues, we use \textsc{PyBDSF} for source-finding and --extraction. In our \textsc{PyBDSF} run, we select an island threshold of 3$\sigma$ and a pixel threshold of 5$\sigma$. Individual Gaussian components are considered to belong to a single source if they occupy the same island. {This assumption is valid for the relatively shallow integration and high resolution of this S-band image.} We enabled the \texttt{adaptive\_rms\_box} to minimise the false detection of artefacts around bright sources and make use of \`{a} trous wavelet decomposition \citep{Holschneider_1989} to improve the fitting of extended sources. We detect 670 sources in the $R=-0.5$ ($4.0"\times 2.4"$) image and 1199 sources in the $R=0.3$ ($6.8"\times 3.6"$) image (Table~\ref{tab:imaging}). Further assessment of the effectiveness of the source-finding method is discussed in Sect.~\ref{sec:completeness}.  

\subsection{Cross-matching with L-band}\label{sec:crossmatch}
To compare our results with L-band, we retrieve the L-band DEEP2 catalogue from \citet{Matthews_2021} and associated L-band image \citep{Mauch_2020} from the SARAO archives (Project ID SCI-20180426-TM-01). 
For a comparison with \citet{Matthews_2021}, we cross-match the L-band and S-band $R=0.3$ catalogues. We select the $R=0.3$ ($6.8"\times 3.6"$) catalogue for this comparison due to the larger number of sources. We match sources by checking whether source components in the L-band catalogue are contained within the S-band sources, where we define the area of each S-band source as an ellipse with major and minor axes equal to twice the FWHM of the source axes. In total, we find 1289 matches with the \citet{Matthews_2021} L-band catalogue, where some S-band sources match to multiple components in the L-band catalogue. This results in matches to 1097 of the S-band sources. With a less conservative cut in the distance from the pointing centre, the S-band catalogue has 92 sources that do not fall within the area of the L-band catalogue. 

\begin{figure}
    \centering
    \includegraphics[width=0.95\columnwidth]{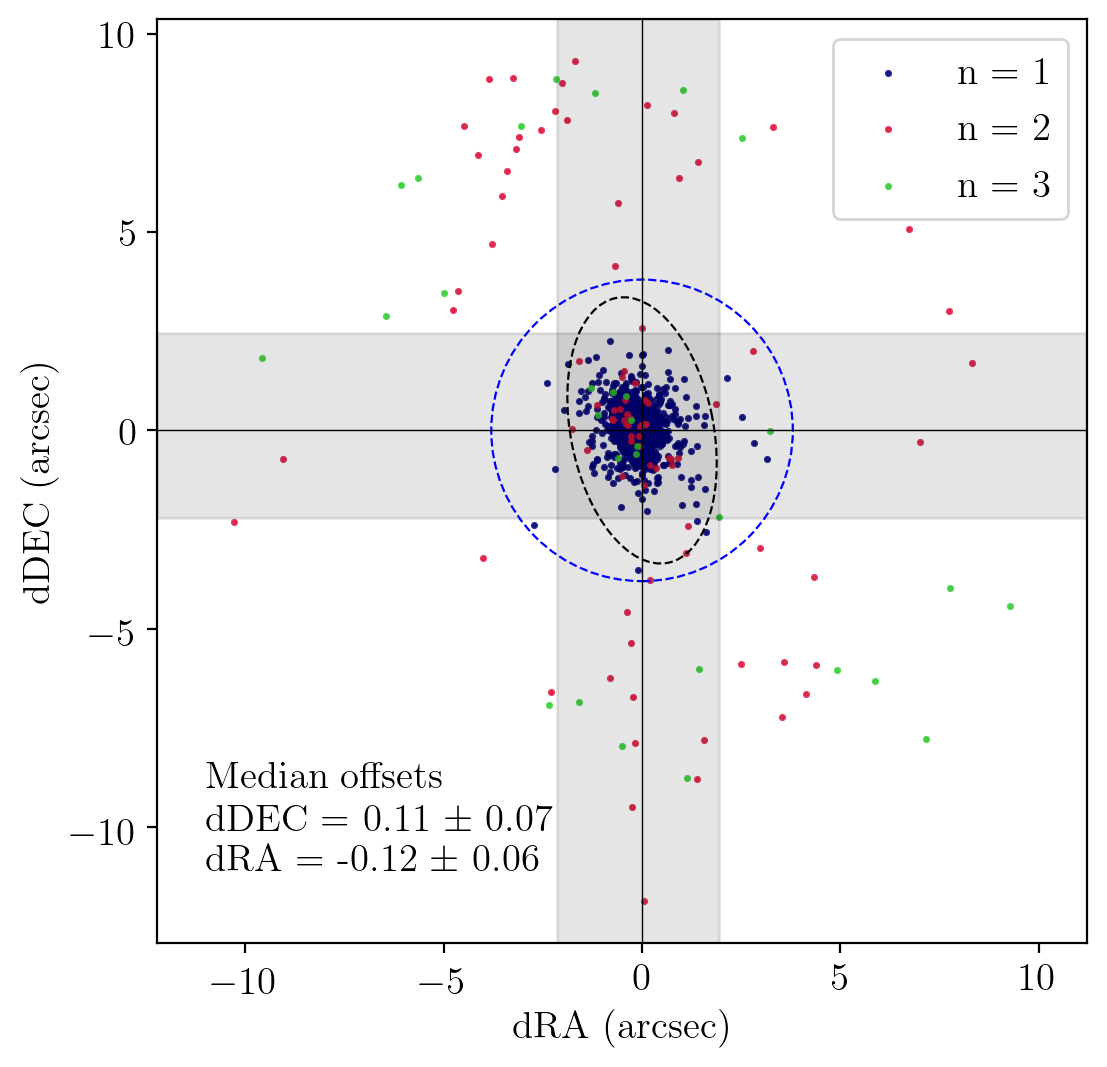}
    \caption{Astrometric offsets of detected sources in the $R = 0.3$ image with respect to the cross-matched components in the L-band observations. The points are colourised according to the number of matches $n$ in the L-band component catalogue for a given S-band source. The synthesised beams of the S- and L-band observations are shown by the black and blue dashed ellipses, respectively. The standard deviations of the offsets are indicated by the shaded grey regions, and the median offsets are listed on the plot. The inset shows the central 10 arcsec of the plot.}
    \label{fig:astrometry}
\end{figure}

To verify the astrometric accuracy of the S-band observations, we measure the positional offsets of the S-band $R = 0.3$ catalogue with respect to the L-band catalogue. The astrometric offsets are plotted in Fig.~\ref{fig:astrometry}, and colourised by the number of matches in the L-band component catalogue for a given S-band source in the $R = 0.3$ image.{ The positions of the L-band and S-band detections show good agreement with small overall offsets of $\Delta \mathrm{RA} = -0.12 \pm 0.06^{\prime\prime}$ and $\Delta \mathrm{DEC} = 0.11 \pm 0.07^{\prime\prime}$. This is well within the pixel size of our images of $0.7^{\prime\prime}$.} The majority of outliers are sources with multiple matches in the L-band, i.e. multi-component sources. We do not correct for these negligible positional offsets in the catalogue.

Given the expected spectral shapes of these sources and the relative noise levels of the respective images, we do not expect sources that are detected in the S-band images to be undetected in the L-band image. We find 10 of our sources in the $R=0.3$ catalogue do not have counterparts in the L-band \citet{Matthews_2021} catalogue. After visual inspection, we find that the the majority of these sources do have associated emission in the \citet{Mauch_2020} image with S/N $ > 5$ and are secondary components to extended sources in L-band. Four sources do not correspond to flux density with S/N $ > 5$ in the L-band image. Two of these sources have associated emission in the L-band image that is below the detection threshold. Using the peak flux density at the position of these sources in the L-band image, we calculated the two-point spectral indices $\alpha$, and find for these sources $\alpha > 2.5$. These extreme values are unphysical \citep{Rees_1967}, and we therefore consider the possibility that these are variable sources. However, because these sources have low S/N, i.e. around the detection threshold, we cannot say with any certainty whether these are variable sources, false detections, or boosted by peaks in the additive Gaussian noise. The remaining two sources do not match any emission in the L-band image. One has a S/N $\sim 5$ at S-band, and is likely a false detection. For the remaining source J041224$-$802217 (S/N = 9.5), a count of one potential variable source out of the total 1199 sources is consistent with the expected variable source ratio for moderately variable extragalactic sources at L-band \citep[$<30$\% variability on $\sim$weekly timescales,][]{Sarbadhicary_2021}. However, we would require broader bandwidth coverage and observations at multiple epochs to confirm the variability of this source. Considering the possibility of this source being a variable radio star, we check for any matches in the Gaia Data Release 3 catalogue \citep{Gaia_2023}. The closest match is a star at RA $=$ 04:12:27.9, Dec $=$ --80:22:18.42, with a 7.4'' separation from our variable candidate. However, the separation is larger than our angular resolution, particularly for the minor axis of the beam, which is aligned with the direction of the source. It is therefore unlikely that this is the same object. 

\subsection{Source properties}

\subsubsection{Multi-band spectral indices}\label{sec:specind}
\begin{figure}
    \centering
    \includegraphics[width=0.9\columnwidth]{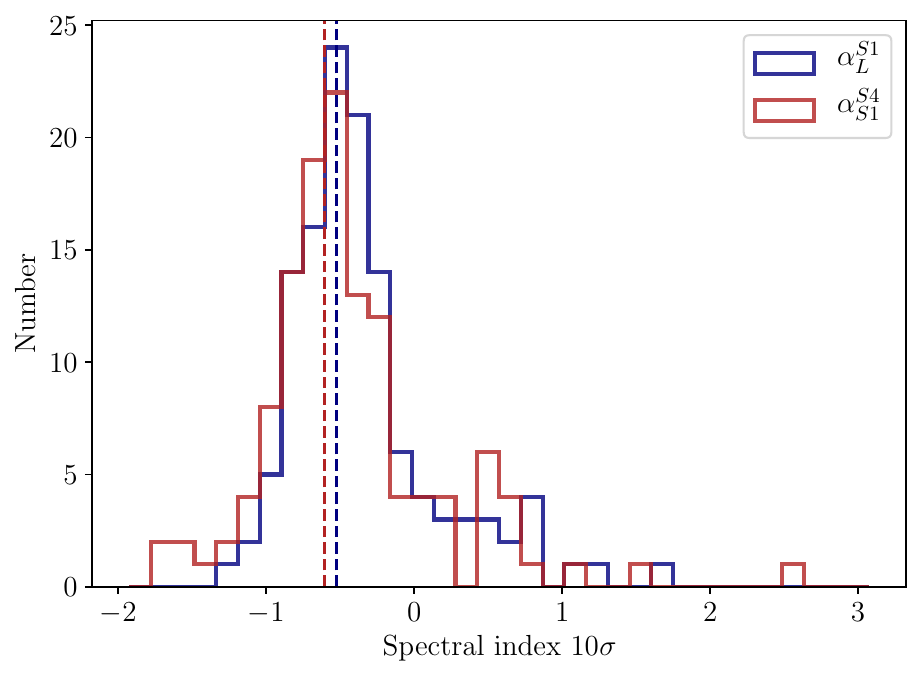}
    \caption{Distribution of spectral indices for $\alpha_\mathrm{L}^\mathrm{S1}$ (blue) and $\alpha_\mathrm{S1}^\mathrm{S4}$ (red) for sources above {$10\sigma$}. The medians of the respective distributions are indicated by the dashed lines.}
    \label{fig:specinddist}
\end{figure}

The spectral energy distributions of radio sources are particularly useful for the classification of sources and the dominant emission mechanism \citep[e.g.][]{Prandoni_2010,Sinha_2023}. Including the L-band observations from \citet{Mauch_2020}, increases our effective bandwidth to $\Delta\nu\sim 1.8$~GHz, allowing for the determination of multi-band spectral indices of the detected sources. As discussed in Section~\ref{sec:calibration}, we make use of 13 SPWs from the S1-band and 6 SPWs from the S4-band. Although the L-band data from \citet{Mauch_2020} was also divided into 14 spectral windows, only the full-band images and catalogues are publicly available, giving us a single flux density value at a frequency of 1.28 GHz for each source \citep{Matthews_2021}. For each source in the $R=0.3$ ($6.8"\times 3.6"$) catalogue, we extract the flux densities (in Jy) within an ellipse with major and minor axes at 1.5 times the source extent. This was done using the \texttt{imfit} task from \textsc{CASA}. We {determine spectral indices for sources above $10\sigma$, where $\sigma$ is the local rms. This is determined from the \texttt{Isl\_rms} output in \textsc{PyBDSF}, scaled to the bandwidth of the individual spectral windows, i.e. by $\sqrt{N_\mathrm{SPW}}$, where $N_\mathrm{SPW}$ is the number of SPWs. For this sample of 125 sources, we calculate two spectral indices. }Firstly, by fitting a power law to the L-band flux density measurement and the 13 SPWs from the S1 band $\alpha_\mathrm{L}^\mathrm{S1}$. Secondly, we fit a power law to the combined {19} SPWs across S1 and S4 $\alpha_\mathrm{S1}^\mathrm{S4}$. The distributions of $\alpha_\mathrm{L}^\mathrm{S1}$ and $\alpha_\mathrm{S1}^\mathrm{S4}$ for the {$10\sigma$ sample } are plotted in Figure~\ref{fig:specinddist}. The spectral index distributions are as expected for the observed frequency range \citep{Condon_1984}. We find a median $\alpha_\mathrm{L}^\mathrm{S1}$ of $-0.52 \pm 0.16$. For $\alpha_\mathrm{S1}^\mathrm{S4}$, we determine a median of $-0.61 \pm 0.33$. The associated errors are determined from the uncertainties on the power-law fits. The distribution of spectral indices including the L-band and that including the S4 sub-band show good agreement with each other. However, a broader bandwidth is required to identify spectral turnovers with significant reliability. While we expect an increase in the number of flat-spectrum sources due to the selection bias at this frequency range, where sources become increasingly core-dominated \citep{Whittam_2013}, both spectral index distributions have a mode of $\alpha = -0.6$. 

\subsubsection{Identifying resolved sources}
\label{sec:resolved}

Due to uncertainties in source fitting, resolved sources cannot be simply identified as sources that are larger than the beam. To confidently identify resolved sources, we consider that due to the aforementioned fitting uncertainties, unresolved sources have Gaussian uncertainties on their measured size parameters. A proxy for the size of a source is the ratio of total flux density to peak flux density, $S_\mathrm{tot}/S_\mathrm{peak}$. The distribution of $S_\mathrm{tot}/S_\mathrm{peak}$ is expected to be log-normal \citep[e.g.][]{Franzen_2015,Wagenveld_2023}, with standard deviation $\boldsymbol{\sigma_\rho}$ described by the sum in quadrature of the relative uncertainties on $S_\mathrm{tot}$ and $S_\mathrm{peak}$,
\begin{equation}
    \sigma_\rho = \sqrt{\left(\frac{\sigma_{S_\mathrm{tot}}}{S_\mathrm{tot}}\right)^2 + \left(\frac{\sigma_{S_\mathrm{peak}}}{S_\mathrm{peak}}\right)^2}.
\end{equation}
The \textsc{PyBDSF} fit uncertainty for $S_\mathrm{tot}$ includes the uncertainty on $S_\mathrm{peak}$ as well as the uncertainty on the determined major and minor axes of the source \citep[Eq. 36 of ][]{Condon_1998}. To make sure $\sigma_{S_\mathrm{tot}}$ and $\sigma_{S_\mathrm{peak}}$ are independent, we subtract the the fitted uncertainty on $S_\mathrm{peak}$, and define $\sigma_{S_\mathrm{peak}}$ as the local rms noise at the source position. We then identify resolved sources as sources for which
\begin{equation}
    \ln \left(\frac{S_\mathrm{tot}}{S_\mathrm{peak}}\right) > 2 \sigma_\rho,
\end{equation} 
which should reliably select 97\% of unresolved sources.

\begin{figure*}
    \centering
    \includegraphics[width=0.8\textwidth]{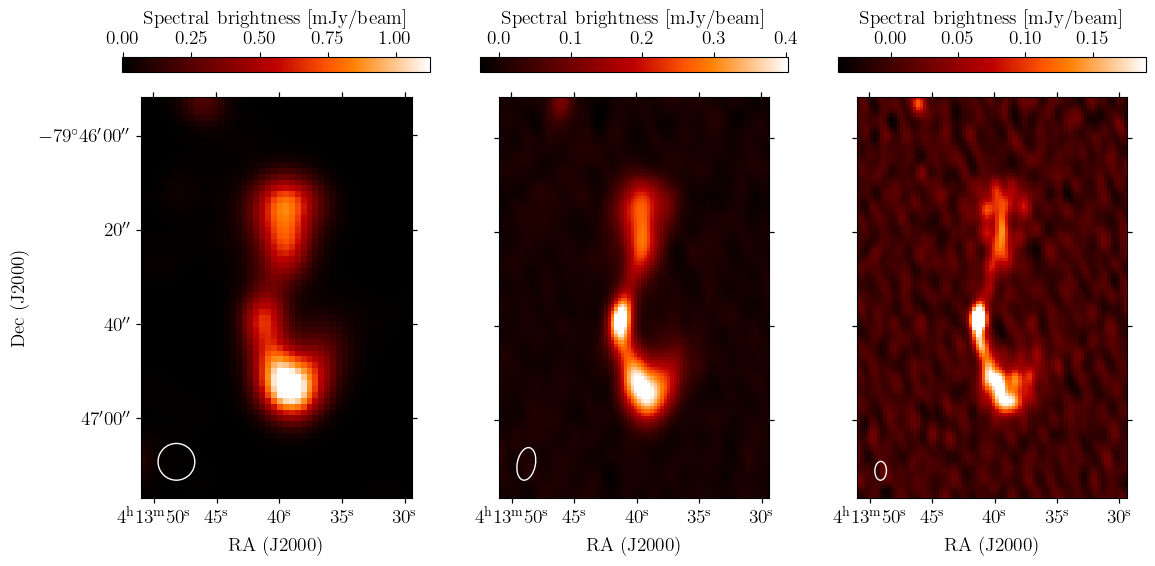}
    \includegraphics[width=0.8\textwidth]{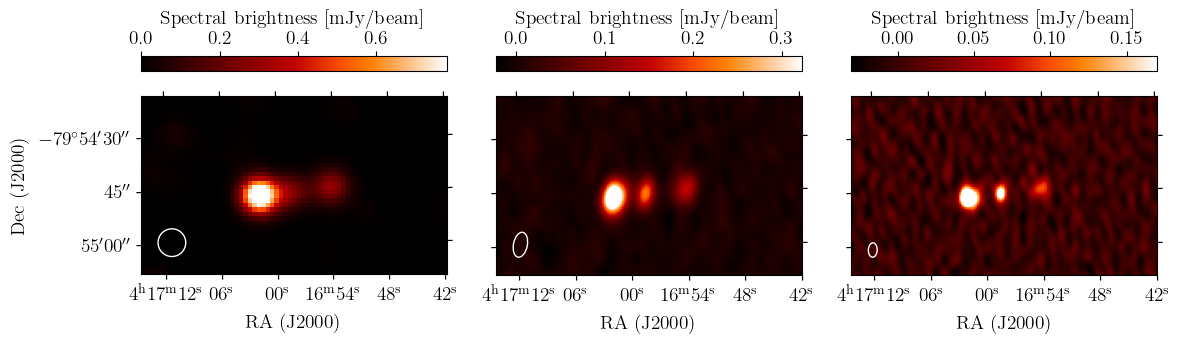}
    \includegraphics[width=0.8\textwidth]{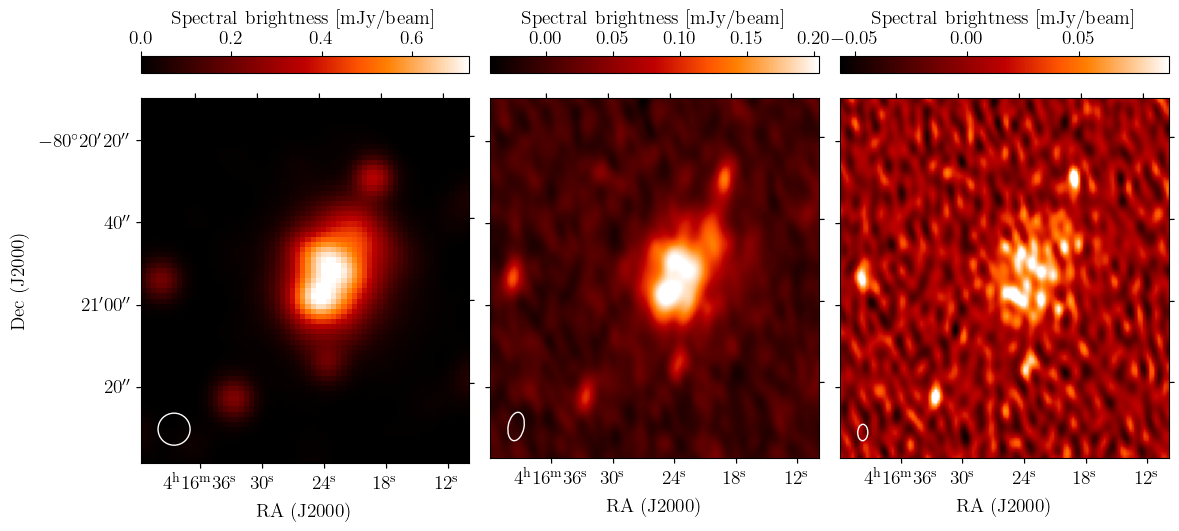}
    \caption{Cutouts of selected extended sources in L-band \citep[][left]{Mauch_2020}, S-band $R=0.3$ (centre) and $R=-0.5$ (right). The restoring beam for each image is shown in the bottom left corner. From top to bottom, the sources shown are J041339--794637, J041654--795445 and J041622--802052.}
    \label{fig:cutouts}
\end{figure*}

Sources are classified as multi-component sources, if they are fitted with multiple Gaussian components with PyBDSF. By this definition, all multi-component sources should also be classified as resolved. With the higher resolution of the S-band $R = -0.5$ ($4.0"\times 2.4"$) image, we find an increased percentage of both resolved and multi-component sources. We find {18\%} resolved sources with $R = 0.3$ and {22\%} resolved sources for $R = -0.5$. Similarly, we find 7\% of sources in $R = 0.3$ are multi-component sources, and 10\% multi-component sources for $R=-0.5$. In the \citet{Matthews_2021} catalogue, there is no classification for resolved sources but they identify 35 multi component sources (0.2 per cent of the total number of sources). At low flux densities ($< 15$ \textmu Jy), we expect the detected samples in L- and S-band to be dominated by star-forming galaxies, which should be mostly unresolved at MeerKAT S-band angular resolution \citep{Cotton_2018}.

Figure~\ref{fig:cutouts} illustrates some examples of the structural detail revealed by S-band observations in resolved sources. We plot cutouts of extended sources at L-band \citep{Mauch_2020}, and S-band $R=0.3$ ($6.8"\times 3.6"$) and $R=-0.5$ ($4.0"\times 2.4"$). The top row is a radio galaxy J041339--794637, for which we resolve the jets in S-band, whereas in the L-band image, only the core and lobes are visible. The $R=-0.5$ image also resolves multiple bright components in the southern lobe, unseen at L-band. The middle row shows an example of a source J041654--795445 that is detected as two components at L-band, but as three components and two discrete sources at S-band. Assuming this is a radio galaxy with a central core and two lobes, the core is associated to the same source as the western lobe (Section~\ref{sec:sourcefinding}). The third row shows star-forming galaxy (SFG) J041622--802052 \citep{Gaia_2020} at a redshift of $z=0.016$ \citep{Fairall_2006}, which is resolved in all images, with extended emission recovered in the $R=0.3$ image (centre). However, we do not fully recover the extended emission at $R=-0.5$ due to the higher angular resolution. This illustrates how with the brightness temperature sensitivity of MeerKAT, it is possible to detect extended SFGs at S-band frequencies, even with relatively short integration times \citep[e.g.][]{Condon_2015}, and shows the great potential for larger area snapshot surveys with the MeerKAT S-band for both AGN and star-forming populations. 

\subsection{Catalogue overview}
\label{sec:catalogue}

\begin{figure}
    \centering
    \includegraphics[width=0.95\columnwidth]{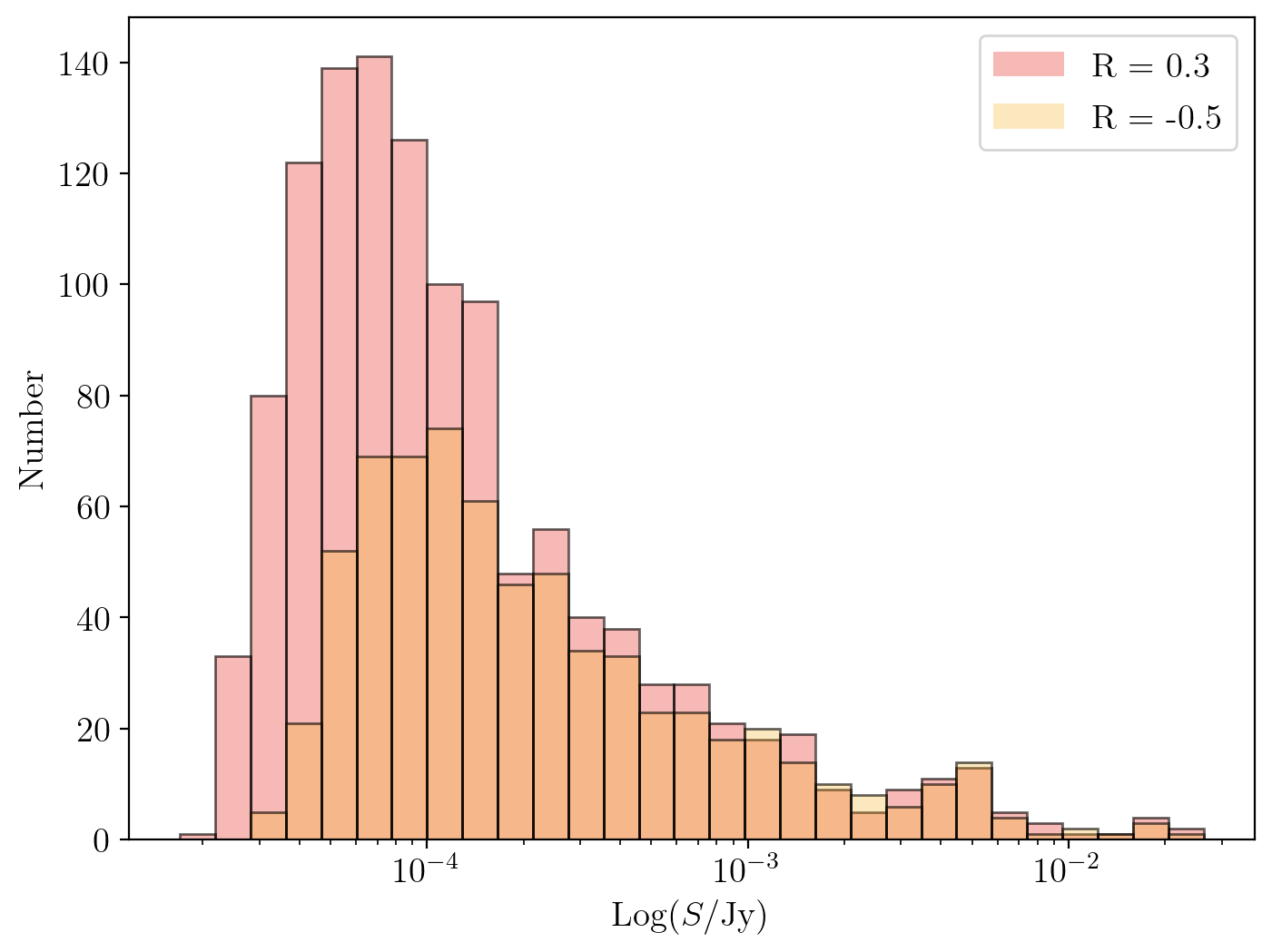}
    \caption{Distribution of source flux densities from the $R=0.3$ ($6.8"\times 3.6"$, red) and $R = -0.5$ ($4.0"\times 2.4"$, yellow) catalogues described in Section~\ref{sec:catalogue}.}
    \label{fig:flux-dist}
\end{figure}

The final source catalogues of the DEEP2 field with MeerKAT S-band includes the 1199 sources detected in the $R=0.3$ ($6.8"\times 3.6"$) image and the 670 sources detected in the $R=-0.5$ ($4.0"\times 2.4"$) image. We provide a separate catalogue for each image. The $R=0.3$ catalogue spans a flux density range from 16.9 \textmu Jy to 26.0 mJy and for $R=-0.5$, the flux densities range from 31.1 \textmu Jy to 25.3 mJy, due to the lower sensitivity of the latter weighting. The flux density distribution of each catalogue is shown in Figure~\ref{fig:flux-dist}. The catalogues are available in FITS format on the SARAO archive at \url{https://doi.org/10.48479/zdyz-8342}. An example of the catalogue structure can be found in Appendix~\ref{sec:cat-eg}. Here, we provide the source positions, flux densities, source sizes and their associated errors as determined from \textsc{PyBDSF}. The catalogue also includes the classification of resolved sources (Section~\ref{sec:resolved}) and the spectral index (Section~\ref{sec:specind}), where available. The cross-matched source ID and flux density at 1.28 GHz from \citet{Matthews_2021} has been included for convenience. In brief, we summarise the catalogue columns below, with a detailed description in Appendix~\ref{sec:cat-eg}.\\
\indent 1.--27. \textsc{PyBDSF} output, as described in the \textsc{PyBDSF} documentation\footnote{\url{https://pybdsf.readthedocs.io/en/latest/}}.\\
\indent 28. Boolean classification for resolved sources\\
\indent 29.--30. Corresponding source index and flux density from the \citet{Matthews_2021} 1.28 GHz catalogue. (Only available for $R=0.3$)\\
\indent 31.--32. Spectral index $\alpha_L^{S1}$ over the L- and S1- bands for sources detected above 10$\sigma$, and associated uncertainties (Only available for $R=0.3$). \\
\indent 33.--34. Spectral index $\alpha_{S1}^{S4}$ over the S1- and S4- bands for sources detected above 10$\sigma$, and associated uncertainties (Only available for $R=0.3$).\\

While the $R=-0.5$ image and associated catalogue is particularly useful for a higher resolution perspective of resolved galaxies, we recommend the $R=0.3$ catalogue for general use. The lower rms yields almost a factor of 2 more detected sources and the lower angular resolution of the images allow for a more consistent comparison with the L-band catalogue, and better recovery of faint, extended emission (Figure~\ref{fig:cutouts}). We therefore only provide the L-band catalogue match and spectral indices for the $R=0.3$ catalogue. Furthermore, we recommend $R=0.3$ as an optimal robust weighting for such broadband extragalactic fields with MeerKAT S-band. For a detailed analysis on the selection of optimal robust weighting, see Appendix~\ref{sec:robust}.

\section{Differential source counts}
\label{sec:sourcecounts}

To compare the catalogues with respect to each other and to other surveys, we calculate source count corrections and differential source counts for both the $R = 0.3$ and $R = -0.5$ catalogues. Further results following deeper observations will presented in future publications.

\subsection{Completeness}
\label{sec:completeness}

\begin{figure}
    \centering
    \includegraphics[width=0.95\columnwidth]{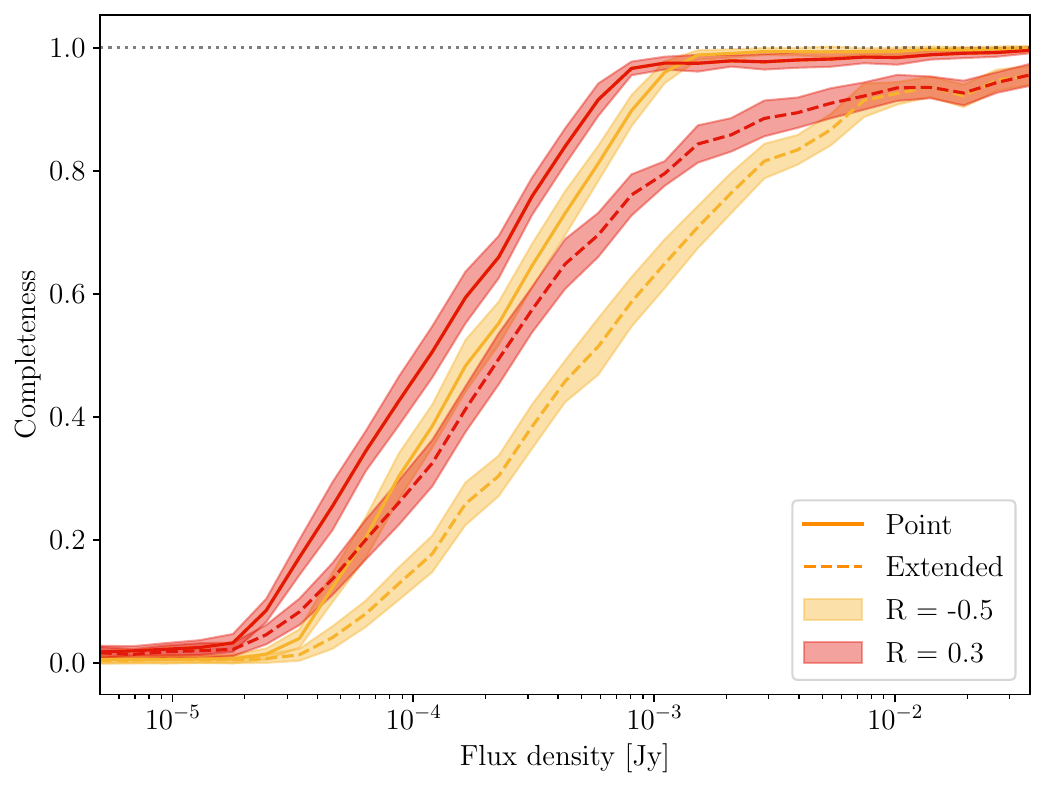}
    \caption{The catalogue completeness for point (solid line) and extended sources (dashed) as a function of flux density. The completeness curves are shown in red and yellow for $R=0.3$ and $R=-0.5$, respectively. The shaded regions show the standard deviation and the black dotted line indicates 100\% completeness.}
    \label{fig:completeness}
\end{figure}

To assess the completeness of our catalogues and the effectiveness of our source-finding strategy, we test our source-finding methods on simulated images. To do this, we make use of the SKA Design Survey (SKADS) Simulated Skies catalogue \citep{Wilman_2008}, {for realistic samples of extragalactic sources}. We create catalogues of 4500 simulated sources, with sources uniformly distributed in logarithmically spaced bins over a flux density range of $ \sigma< S < S_\mathrm{max}$, where $S$ is the flux density, i.e. $n(S) \propto S^0$. The lower limit is equal to the rms $\sigma$ of an image and the upper limit is consistent with the brightest detected source. While this is not a physically realistic flux density distribution, it is the best distribution for determining the completeness with the same level of accuracy across the probed flux density range. We convolve the simulated sources to the angular resolution of our observations and inject them at randomly selected positions in the residual image from \textsc{PyBDSF}. This residual image contains no detectable sources with our chosen source-finding parameters and has consistent noise characteristics with our original image. We then perform the same source-finding routine on the simulated image (i.e. the residual image with injected synthetic sources), with the rms and mean maps determined by \textsc{PyBDSF} from the original image. We consider a source successfully recovered if it is detected within the FWHM of the restoring beam of the injected position. This process is repeated 50 times each for resolved and unresolved sources to achieve a statistically robust result. In the case of resolved sources, we randomly select sources from the SKADS catalogue with major and minor axes larger than zero, and for unresolved sources we randomly select SKADS sources with a major axis of zero, as they are defined as delta functions in the catalogue. Resolved and unresolved sources are respectively modelled as delta functions and Gaussians. These sources are then convolved with the clean beam and injected into the mock images. We calculate the completeness as the ratio between the number of recovered sources to the total number of sources within each flux density bin. The completeness curves for both the $R = -0.5$ and $R = 0.3$ images are plotted in Fig.~\ref{fig:completeness}. The respective completeness curves for resolved and unresolved sources do not follow the same distribution, as expected due to the reduced surface brightness of resolved sources. We find that 98.5\% of point sources are recovered at flux densities above $5\times 10^{-5}$~Jy for $R = 0.3$ and $1\times 10^{-4}$~Jy for $R = -0.5$. For extended sources, we reach 93\% completeness at ${\sim}10^{-2}$~Jy for both images.

\subsection{Purity}

We assess the false detection rate, or purity of our source finding parameters, to be taken into account when measuring number counts. We do not expect potential false detections to be caused by calibration artefacts around bright sources, as this field was selected in particular not to have any problematic bright sources, with the brightest source in the field at 26~mJy. Any false detections are likely attributed to statistical outliers in the noise, such that the number of negative detections is representative of the number of positive false detections. To determine the purity, we invert the pixel values of the images and execute PyBDSF using the rms and inverted mean maps from the original images, to ensure that the source finding is performed in the same way. In both images, we find zero false detections above 5$\sigma$ which are attributed to outliers in the noise, i.e. a catalogue purity of 100 per cent. As a further assessment for the false detection rate, we assess the Gaussianity of the noise, by plotting the flux density distributions of the four corners (300 pix $\times$ 300 pix) of the primary beam-uncorrected image ($R=0.3$). We perform Komolgorov-Smirnov tests on these distributions, resulting in an average p-value of 0.71. As such, we do not reject the null hypothesis that the image noise is Gaussian, and we do expect expect extreme outliers in the noise that could result in a false detection. However, as per the analysis done in Section~\ref{sec:crossmatch}, it is likely that 1--3 of the sources not detected in L-band are false detections.

\subsection{Stacking}
\label{sec:stacking}

To obtain a flux density estimate for the sources below our detection threshold, we make use of image-plane stacking. {Stacking can be used to statistically determine the average flux density properties of undetected source samples by co-adding their emission, based on their known positions from other catalogues \citep[e.g.][]{Dunne_2009,Karim_2011,Perger_2024}}. To do this, we select all sources from the L-band catalogue that were not detected in the $R=0.3$ ($6.8"\times 3.6"$) S-band image ($<5\sigma$) and lie within 0.4$^{\circ}$ of the S-band phase centre (i.e., close to the primary beam FWHM). This resulted in a {stacking }sample of 7573 sources. We generated 100$\times$100 pixel cutouts of all sources in the stacking sample{, such that the source position in our L-band catalogue was centred in the cutout and smoothed the images to the angular resolution of the L-band image (7.6''$\times$7.6''). We calculate the stacked mean flux density over these cutouts. Although the median stacking is particularly useful as it is not sensitive to outliers, its interpretation is limited for faint sources \citep{White_2007}, for which the mean flux density is more appropriate. We can account for the local noise for each source, by taking a weighted mean for the stacked flux density, 
\begin{equation}
    S_\mathrm{mean} = \frac{\sum _{i=1}^N w_i \times S_i}{\sum _{i=1}^N w_i },
\end{equation}
and constructing a mean map. Here, $S$ is the pixel flux density, $w$ is the weight, as function of the local noise $\sigma_\mathrm{local}$, determined as $w = \sigma_\mathrm{local}^{-2}$, and $N$ is the number of stacked sources \citep[e.g.][]{Zwart_2014}. We determine the stacked flux density by fitting a 2D Gaussian to the stacked image. The associated uncertainties are calculated as the rms over all stacked subimages divided by $\sqrt{N}$.

We stacked the full sample and make a $62\sigma$ weighted stacked detection, with a flux density of $10.40 \pm 0.17$~\textmu Jy. We aim to use the stacked flux density estimates to probe the source counts below the detection threshold. For this, we divide the sample into three logarithmically spaced flux density bins, using the flux density measurement from the \citet{Matthews_2021} catalogue. The number of sources per bin and the resulting stacked flux density is summarised in Table~\ref{tab:stacking}. With a large number of sources in each bin, we can successfully make high S/N stacked detections of the faint source population in the DEEP2 S-band $R=0.3$ image. However, when measuring the mean flux density of the stacked sources at L-band and scaling to 2.5 GHz (assuming an average spectral index of $\alpha = -0.7$), we find that $S_\mathrm{mean}$ is, on average, 26\% lower than expected. As this would imply that the faint source population have steeper spectral indices of $\alpha = -1.2$, we consider that this missing flux density is likely a systematic effect.
\begin{table}
    \centering
    \caption{Stacked flux density and number of stacked sources for logarithmically spaced bins.}
    \begin{tabular}{ccc}
    \hline
      Bin width [\textmu Jy] & N & $S_\mathrm{mean}$ [\textmu Jy] (S/N) \\ 
      \hline
      $10< S_\mathrm{1.28 GHz} <16$ & 2666 & $5.54\pm0.17$ (33) \\
      $16< S_\mathrm{1.28 GHz} <25$ & 2211 & $8.47\pm0.18$ (45) \\
      $25< S_\mathrm{1.28 GHz} <40$ & 1454 & $14.00\pm0.23$ (60) \\
      \hline
    \end{tabular}
    \label{tab:stacking}
\end{table}

One such systematic, as reported in \citet{Matthews_2021}, is the offset between the true flux density and the measured flux density in the DEEP2 L-band for sources with $S_\mathrm{1.28 GHz} \lesssim 40$~\textmu Jy. This offset is attributed to source confusion at low flux densities. For the flux densities in the stacked source sample, the distribution of these offsets has a positive tail that ranges up to $\sim 10$~\textmu Jy. While this contributes to the inconsistent spectral index measured here, it likely doesn't account for the full discrepancy. There are likely additional errors introduced by the limitations of stacking which may lead to an underestimation of the S-band flux density. For example, sources below the noise are not deconvolved \citep{Karim_2011}, which will result in a loss of signal in the main lobe of the PSF. Future deep observations will allow for a more accurate determination of the spectral indices for the \textmu Jy 3~GHz source population of the DEEP2 field.

\begin{figure*}
    \centering
    \includegraphics[width=0.75\textwidth]{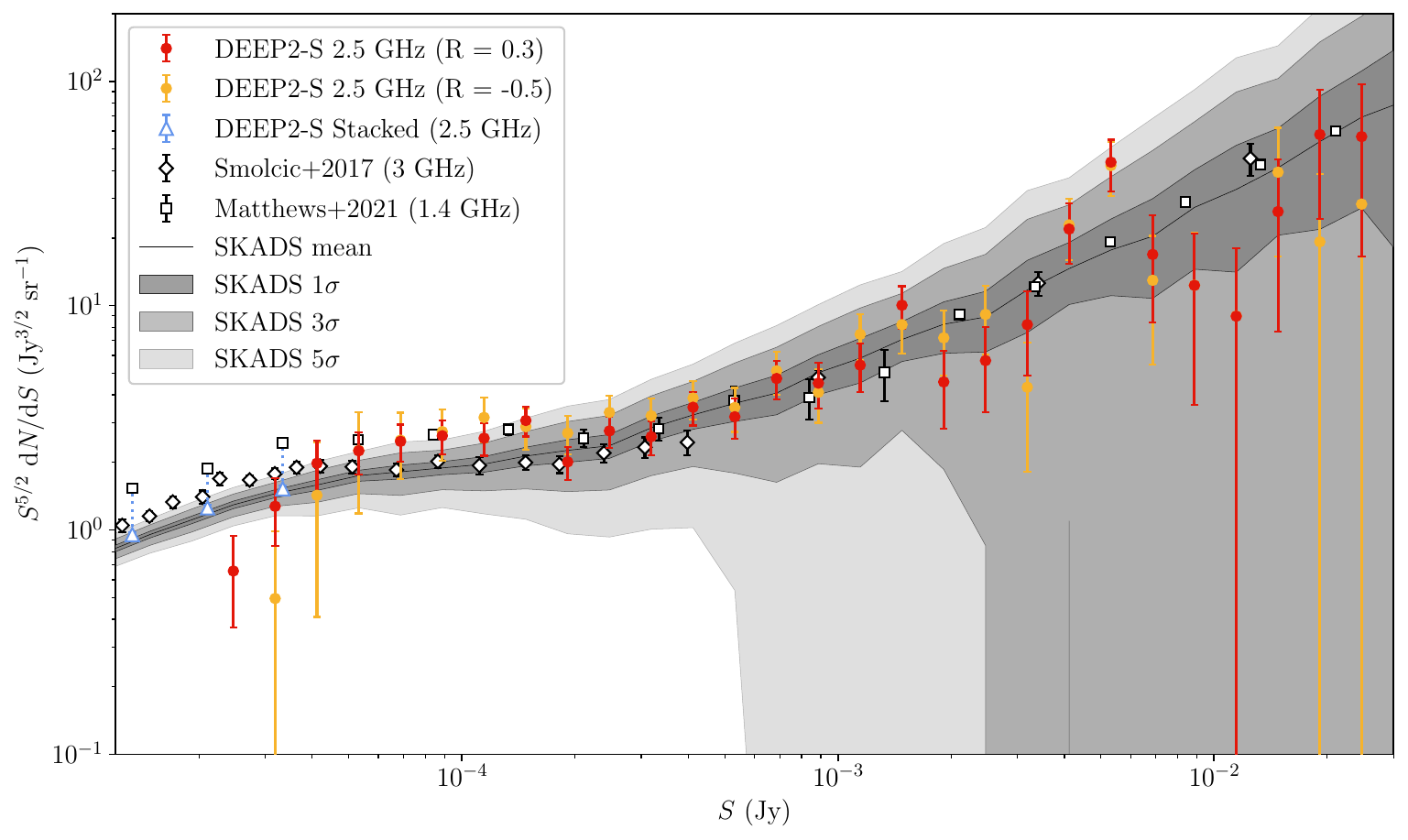}
    \caption{Completeness-corrected differential source counts for the DEEP2 S-band catalogues with $R = 0.3$ (red) and $R = -0.5$ (yellow). The expected differential source counts from the SKADS catalogue is shown by the black solid line, with the grey shaded regions indicating the simulated 1-, 3- and 5$\sigma$ variance stemming from our limited sky coverage and Poisson variance. The 1.4~GHz source counts of DEEP2 and NVSS \citep{Matthews_2021} and 3~GHz VLA COSMOS source counts \citep{Smolcic_2017} are shown as white squares and diamonds respectively, converted to 2.5~GHz, assuming a spectral index of $\alpha=-0.7$. For the majority of the probed flux density range, source counts agree between each other. In the range where where both simulations and different observations begin to disagree ($S \lesssim 100$~\textmu Jy) our counts are still broadly consistent with \citet{Matthews_2021} before they fall off due to increasing incompleteness at around 50~\textmu Jy. The stacked source counts (blue triangles) are computed assuming that all missing flux density in the stacking is due to missing sources, presenting a lower limit on S-band source counts below our current detection limit, which is consistent with the expected source counts.}
    \label{fig:sourcecounts}
\end{figure*} 

\subsection{Euclidean number counts}
 
The differential source counts of these catalogues are shown in Fig.~\ref{fig:sourcecounts}, and are also tabulated in Appendix~\ref{sec:sourcecountsapp}. The source counts are corrected for completeness, assuming that point sources constitute a third of the population, with the rest being extended\footnote{Note that extended here simply means a major axis larger than zero in the SKADS catalogue with which the completeness simulations were performed. With this definition, star-forming galaxies are always extended even though most are likely unresolved with the current image resolution \citep{Cotton_2018}.}. This follows the relative abundances of intrinsic source sizes in the SKADS catalogue, and thus should represent the real size distribution of sources to the extent that SKADS sources accurately represent the observed source population. The uncertainty in the completeness (shown in Fig.~\ref{fig:completeness}) is also accounted for in the source count uncertainties. Aside from uncertainties in source counts that arise from Poisson statistics {and completeness}, cosmic variance in source counts in a limited sky area is also expected to be affected by source clustering. Following \citet{Heywood_2013}, we quantify these effects on differential source counts using the SKADS catalogue \citep{Wilman_2008}. From the SKADS catalogue, we randomly extract sky areas with an area equal to that of the images, at a frequency of 2.5~GHz (matching the mean frequency of the combined S1 and S4 images), and compute the differential number counts. We do this 500 times to statistically determine the expected variance in source counts, both due to clustering and Poisson contributions. This variance is also shown in Fig.~\ref{fig:sourcecounts}, illustrating that down to the completeness limit, the variation in differential source counts of the two S-band images can be explained by the expected variance.

As a comparison for the source counts obtained here, Fig.~\ref{fig:sourcecounts} includes the combined 1.4~GHz DEEP2 and NVSS source counts from \citet{Matthews_2021}, as well as the 3~GHz VLA-COSMOS source counts from \citet{Smolcic_2017}. Both have been converted to 2.5~GHz assuming a spectral index of $\alpha=-0.7$. At higher flux densities, all measurements agree well within the variance introduced by the small sky coverage of DEEP2. However, at lower flux densities ($S \lesssim 100$~\textmu Jy), we enter a regime where different measurements as well as the theoretical expectation start to significantly disagree. Taking into account their respective sky coverages, both the \citet{Matthews_2021} and \citet{Smolcic_2017} counts are larger than the expected source counts from SKADS. This is commonly attributed to an underestimation of SFGs in the SKADS catalogue \citep[e.g.][]{Smolcic_2017,Vandervlugt_2021,Hale_2023}. Though our S-band source counts are roughly in agreement with {the \citet{Matthews_2021} source counts}, our catalogue is highly incomplete at these flux densities (see Fig.~\ref{fig:completeness}), {resulting in large uncertainties. At roughly 50~\textmu Jy the source counts fall off, for which the completeness corrections are no longer adequate.}  

As demonstrated in Section~\ref{sec:stacking}, we can use image plane stacking to detect sources below the noise using the source positions from the L-band catalogue. We can also use this to probe the source counts below the detection threshold. We do this by stacking at the positions of L-band sources for each flux density bin of the \citet{Matthews_2021} source counts (Table~\ref{tab:stacking}). Since we can not know how many sources are actually detected in the stacked measurement at S-band, these results can not directly be translated to source counts. However, in measuring the mean flux density of the stacked sample, we can compare it to the expected mean flux density of the sample, assuming, as before, a spectral index of $\alpha=-0.7$. We see that the mean flux densities of the stacked samples are generally low, which could be caused by a number of effects, including systematic effects (as discussed in Section~\ref{sec:stacking}), or a portion of sources not being detected. If we assume that all missing flux density is caused by missing sources, we can convert this to an estimate of source counts. This is shown in Fig.~\ref{fig:sourcecounts} and compared to the \citet{Matthews_2021} source counts, as these represent the source counts expected if all L-band sources are detected in S-band. As it is unlikely that all missing flux density is due to missing sources, these stacked source counts can be viewed as a lower limit on source counts at S-band. It is worth noting that the amount of missing flux density is essentially the same in all bins, and thus not dependent on flux density.

While deeper observations are needed for more robust source counts, the observing time required to reach the necessary depth at S-band is only a fraction of the best available deep S-band surveys. For example, $\sim$5\% of the total integration time of VLA-COSMOS 3 GHz \citep{Smolcic_2016} survey is required to achieve the same point-source sensitivity, as well as improving brightness temperature sensitivity (for S1 sub-band), highlighting the potential for deep S-band surveys with MeerKAT \citep[e.g.][]{Jarvis_2016}. To achieve the sensitivity levels of the deepest published S-band survey, COSMOS-XS \citep{Vandervlugt_2021}, with a sensitivity of 0.53 \textmu Jy beam$^{-1}$, we would require a comparable integration time to COSMOS-XS. However, the MeerKAT S-band returns a 16 times larger sky coverage in a single pointing. In particular, and as discussed above, a larger area has a clear advantage in managing the cosmic variance in differential source counts. In comparison to COSMOS-XS, a survey of similar depth with MeerKAT S-band will reduce the scatter due to source clustering by a factor of 1.7 \citep{Heywood_2013} and Poisson statistics by a factor of 4. The observations could be further optimised by observing in sub-band S2 (Table~\ref{tab:subbands}), which is relatively free of RFI and beam squint effects \citep{deVilliers_2023}, resulting in a larger usable bandwidth. 

\section{Discussion and conclusion}\label{sec:concl}

We have presented pilot S-band catalogues of the DEEP2 field, a unique radio selected legacy field. With images combining the S1 and S4 subbands, we achieved a sensitivity of 4.7 and 7.5 \textmu Jy beam$^{-1}$, for robust values of $R=0.3$ ($6.8"\times 3.6"$) and $R=-0.5$ ($4.0"\times 2.4"$), respectively. We detected 1199 and 670 sources at the respective robust weightings.

While we have found that imaging the combination of two frequency sub-bands ($\sim$1750 MHz) is possible with the MeerKAT S-band system, the higher end of the S4 subband $\nu > 3.06$~GHz has a significantly higher rms and was excluded from MFS images for the best wide-field imaging result. We conducted a spectral index analysis through spectral fitting of the 19 SPWs across the S1 and S4 sub-bands and the measurement at 1.28~GHz from the \citet{Mauch_2020} images, and found a spectral index distribution as expected in the literature. We computed the completeness-corrected differential sources counts for the DEEP2 field at 2.5~GHz down to 20~\textmu Jy, and found our source counts to be consistent with simulations and other deep field source counts. Although these observations do not obtain the sensitivities of the deepest S-band surveys in the literature, only a fraction of the observing time was needed to reach similar depth and achieve comparable results. To probe the faint source population $S<40$~\textmu Jy, we performed image-plane stacking on the $R=0.3$ image and estimated lower limits for source counts down to $10$~\textmu Jy, which we found to be consistent with simulations but lower than other source counts in the literature. We attributed the lower stacked flux density with respect to \citet{Matthews_2021} to missing sources or systematic effects, which may be caused by assumptions made in the stacking procedure.

Future S-band observations of the DEEP2 field, will overcome the confusion limit at 1.28 GHz, allowing for the direct imaging of SFGs below 0.55 \textmu Jy beam$^{-1}$. Furthermore, the availability of a confusion-limited L-band image provides a comparison to a near-complete sky model with the same instrument. Through the computation of differential source counts, we aim to address the completeness for shallow observations with the MeerKAT S-band. This serves to inform future moderate to large scale surveys with short per-target integration times. The addition of the S-band to the MeerKAT system will allow for more detailed studies of extragalactic fields across a broader bandwidths and into the sub-\textmu Jy regime with longer integration times, among other continuum science goals (e.g. Galactic plane, nearby galaxies), providing great scientific potential for future S-band surveys in the pre-SKA era, with MeerKAT and its upcoming 14-antenna extension, MeerKAT+.

\section*{Acknowledgements}

We thank the referee, Allison Matthews, for helpful comments and suggestions that significantly improved this work. We thank Bill Cotton and Rainer Beck for helpful comments on the manuscript. SR and JDW acknowledge the support from the International Max Planck Research School (IMPRS) for Astronomy and Astrophysics at the Universities of Bonn and Cologne. RPD acknowledges funding from the South African Research Chairs Initiative of the Department of Science and Innovation and National Research Foundation (Grant ID 77948). M. R. Rugel is a Jansky Fellow of the National Radio Astronomy Observatory. The MeerKAT telescope is operated by the South African Radio Astronomy Observatory, which is a facility of the National Research Foundation, an agency of the Department of Science and Innovation.


This work has made use of the “MPIfR S-band receiver system” designed, constructed and maintained by funding of the MPI für Radioastronomie and the Max-Planck-Society. In particular, we acknowledge the contribution and efforts of the experts of the “Electronics" and “Digital Signal Processing" departments of the MPIfR in this project.

This research has made use of the NASA/IPAC Extragalactic Database, which is funded by the National Aeronautics and Space Administration and operated by the California Institute of Technology.

\section*{Data Availability}

All data products specified in the paper are publicly available in FITS format. The MFS images and catalogues for both $R = 0.3$ and $R=-0.5$ can we found on the SARAO archive at \url{https://doi.org/10.48479/zdyz-8342}. Please see Section~\ref{sec:catalogue} and Appendix~\ref{sec:cat-eg} for more details. Raw data or other data products can be made available upon reasonable request to the authors.



\bibliographystyle{mnras}
\bibliography{example} 




\appendix
\section{MeerKAT S-band system}\label{sec:sband}

\begin{figure*}
  \centering
  \includegraphics[width=0.75\textwidth]{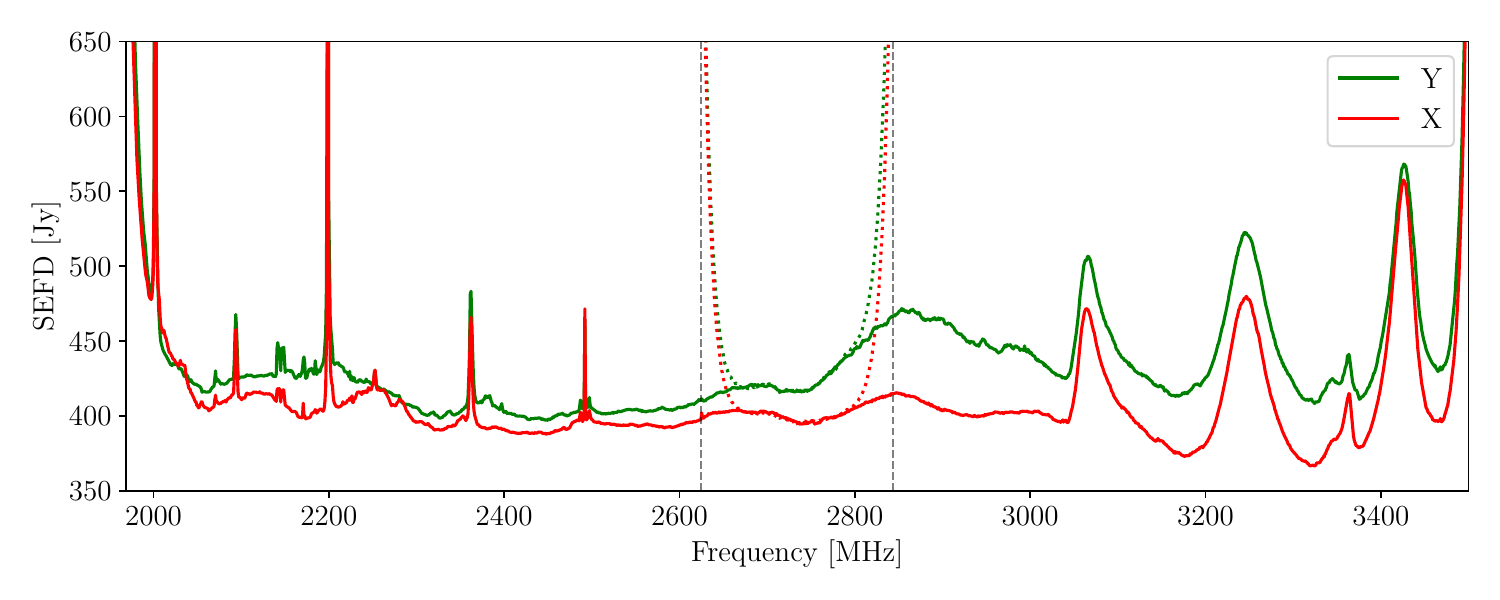}
  \caption{SEFD for a representative antenna over the combined range of our observations (S1 and S4) for both linear polarisations Y (green) and X (red). The band edges are denoted by vertical dashed lines. In the overlapping regions, the nearer sub-band is plotted in solid lines, the other is shown as dotted lines and shows the expected increase towards the edge. The variations in the upper half of the band are caused by frequency dependent gain and noise levels of the receiver system, which cannot be avoided at such a wide band \citep[e.g.][]{deVilliers_2023}. The variations in the gain are stable over time and are removed by the bandpass calibration. Sharp peaks at and below 2500~MHz are caused by various RFI sources, not all of which have been identified yet. }
  \label{fig:sefd}
\end{figure*}

The S-Band receivers \citep{Kramer_2016} were designed and built by the Max Planck Institute for Radio Astronomy (MPIfR), and through a collaboration with SARAO, are now fully integrated into the MeerKAT system, with a receiver on each of the 64 antennas. Their analogue front end and digitisers deliver a full band from 1750 to 3500~MHz. Half of this bandwidth (875 MHz) can be selected with a digital filter at any given time, positioned at five available sub-bands S0--S4 with uniform spacing and large overlap, as summarised in Table~\ref{tab:subbands}. For the observations described in Sect.~\ref{sec:obs}, sub-bands S1 and S4 were used (Table~\ref{tab:bands}). When combined, they provide contiguous coverage of the frequency range 1968--3500~MHz. Within this frequency range, we expect negligible ($< -50$\,dB) frequency aliasing due to the decimation filters. At the very high end, additional sampling-induced aliasing is expected from other Nyquist zones. This is minimised with analogue filters.

\begin{table}
    \centering
    \caption{Frequency range of the MeerKAT S-band sub-bands.}
    \begin{tabular}{cc}
    \hline
     Sub-band   & Frequency range (MHz) \\
     \hline
      S0   & 1750.00 -- 2625.00 \\
      S1    & 1968.75 -- 2843.75 \\
      S2    & 2187.50 -- 3062.50 \\
      S3    & 2406.25 -- 3281.25 \\
      S4    & 2625.00 -- 3500.00 \\
      \hline
    \end{tabular}
    \label{tab:subbands}
\end{table}

In order to measure the absolute flux sensitivity, the noise level per frequency and baseline were determined from the final scans of the flux calibrator J0408--6545 in each of our DEEP2 observations (Section~\ref{sec:obs}) at elevations of 55 and 48 deg in S1 and S4, respectively. These scans were calibrated for the bandpass response, fringe-fitted for dispersive and non-dispersive delays plus residual phases, and the resulting solutions applied. The source spectrum in instrument units was determined from the amplitudes of the calibration solutions. The noise levels in the same units were derived from the data weights, which are based on the auto-correlation powers. To test consistency and confirm the correct scaling, we compared with alternative measures using the rms variations of the visibilities. This was done on the absolute values, but also on differences between adjacent times or frequency channels, so that residual gain variations do not affect the results. All these estimates were consistent with each other.

The baseline-based noise levels were reduced to station-based values and converted to a physical scale using the following flux density model for the calibrator:\footnote{The model was obtained from the \href{https://skaafrica.atlassian.net/wiki/spaces/ESDKB/pages/1481408634/Flux+and+bandpass+calibration\#Using-CASA-setjy-for-non-standard-flux-models}{MeerKAT Knowledge Base}.}
\begin{align}
  \log_{10} S = &-0.9790 + 3.3662\log_{10} f \\
        &- 1.1216\log_{10}^2 f +0.0861 \log_{10}^3 f, \nonumber
\end{align}
where $S$ is the flux density in Jy and $f$ the frequency in MHz. These total noise levels are the sum of the source flux density and system equivalent flux density (SEFD). The resulting SEFD for a representative antenna  is shown as a function of frequency in Figure~\ref{fig:sefd}.

\section{Robust parameters}
\label{sec:robust}

We investigate the optimal Briggs robust parameter to maximise the sensitivity of the observations, the number of sources detected and the angular resolution of our images. To test this, we imaged the concatenated datasets at Briggs robust values from $R = [-2, 2]$, with an interval of 0.5, and a decreased interval of 0.1 between $R=-0.5$ and $R=0.5$. Fig.~\ref{fig:bmaj-robust} shows the resultant image sensitivity and angular resolution of the MFS images as a function of robust weighting. Here, the sensitivity is estimated from the residual images provided by {\tt WSClean}. We note that the rms increases slightly at $R > 0.5$ for the S1 and S4 sub-bands. This trend is not present in the concatenated dataset.
\begin{figure}
    \centering
    \includegraphics[width=0.95\columnwidth]{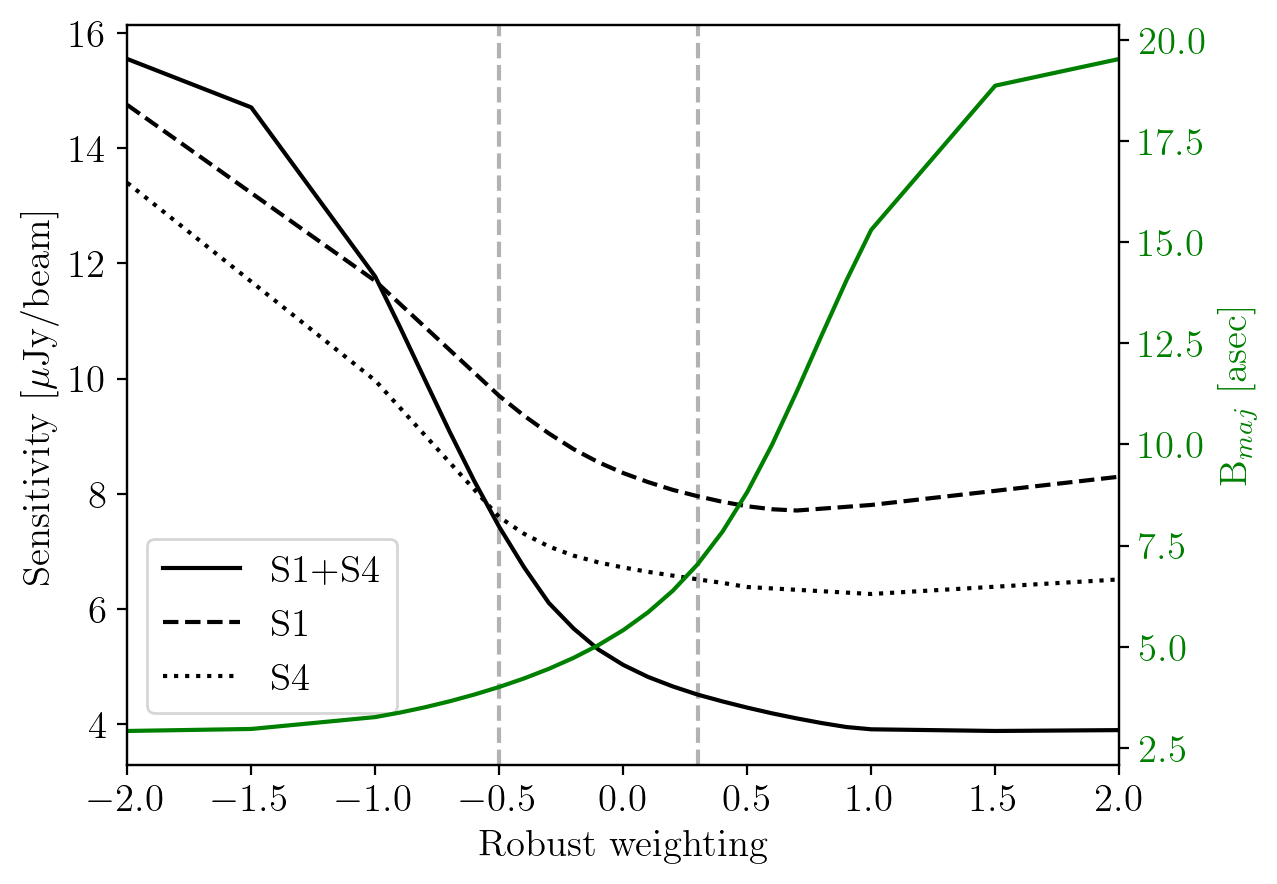}
    \caption{Image sensitivity (black) and the major axis of the restoring beam (green) as a function of robust weighting. The concatenated data is indicated by the solid lines, and the S1 and S4 sub-bands are shown as dashed and dotted lines, respectively. The grey dashed {vertical} lines indicated the chosen weighting values for the final images.}
    \label{fig:bmaj-robust}
\end{figure}
For completeness, we execute source-finding for all images with robust weighting [$-$2, 2], as above. In Fig.~\ref{fig:nsource-rob}, we show the number of detected sources, as a function of robust weighting. As expected, the number of sources increases with higher sensitivity, with the maximum number of sources detected at $R=0.8$. {At $R=0.8$ we have an angular resolution of $\theta = 13^{\prime\prime}$ at a sensitivity of $\sigma = 4.01$~\textmu Jy~beam$^{-1}$, which approaches the theoretical limit for the thermal noise of $\sigma = 3.38$~\textmu Jy~beam$^{-1}$. The thermal noise is estimated for an observation using 55 antennas, and for the bandwidth and flagging percentage in the images presented here, and the image sensitivity is calculated as specified in Section~\ref{sec:calibration} for Fig.~\ref{fig:bmaj-robust}.} At higher robust values, the number of detected sources decreases, as source confusion and PSF sidelobes start to add to the noise. {To quantify this, we consider the residuals between the normalised PSF and restoring beams. We compute this for the cross-section across the major and minor axes of the respective beams, and take the standard deviation $\sigma_\mathrm{PSF,res}$ to quantify the residuals. The cross-sections of the PSF across the major and minor axes for each robust value are shown in Fig.~\ref{fig:beamsize}. Between $-2.0 < R < 0.5$, we have typical $\sigma_\mathrm{PSF,res}$ between 0.02 and 0.04, whereas we find increased residuals beyond $R=0.6$, rising to $\sigma_\mathrm{PSF,res} = 0.09$ for $R = 2.0$. The sidelobe amplitudes amount to $\sim$20\% for $R=0.8$, and increase to $\sim$40\% for $R=2.0$. Images with such weightings for for short observations with MeerKAT S-band will be limited in their scientific interpretation. These weightings should be avoided due to resulting systematic errors \citep[e.g.][]{Radcliffe_2024}. 
\begin{figure}
    \centering
    \includegraphics[width=0.95\columnwidth]{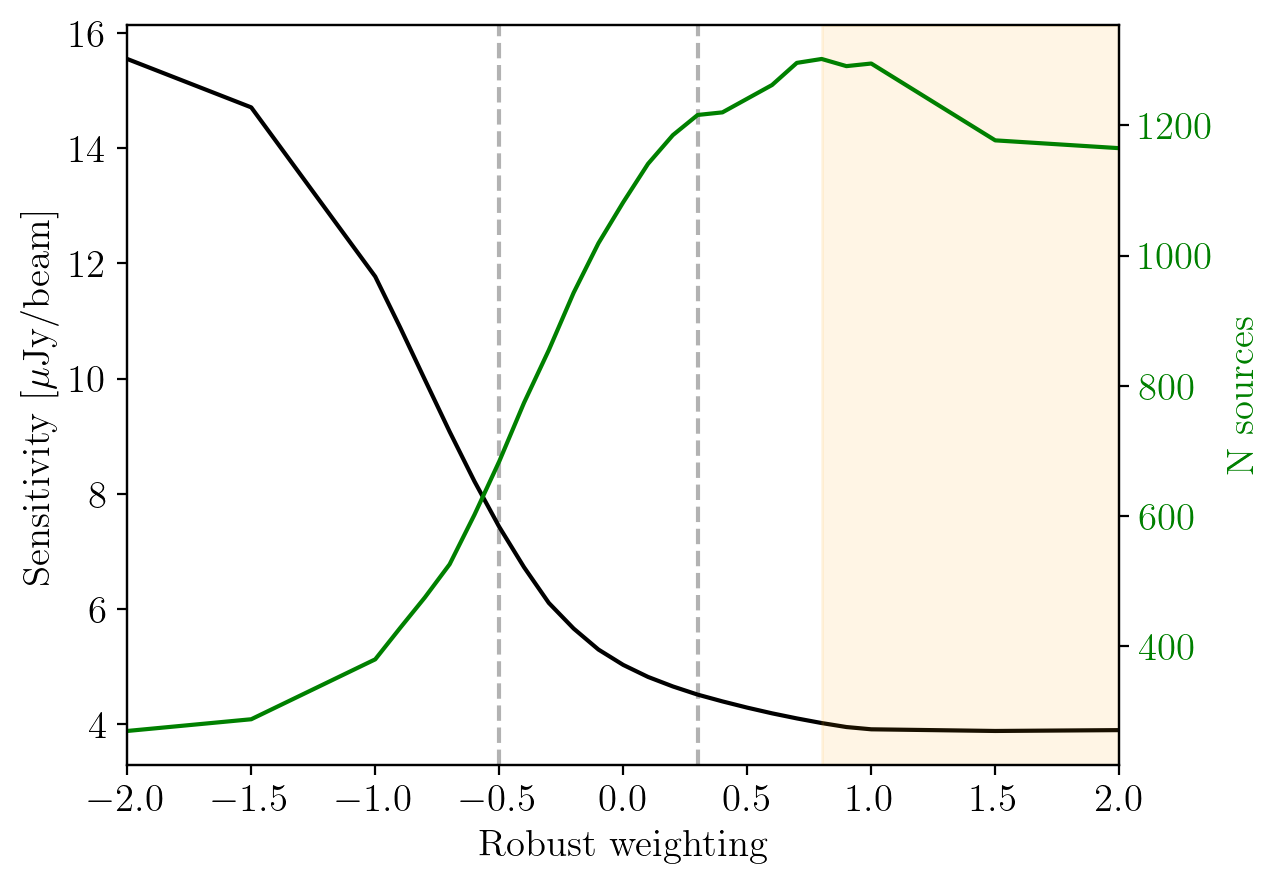}
    \caption{Image sensitivity (black) and the number of detected sources (green) as a function of robust weighting for the concatenated sub-bands. The grey dashed lines indicated the chosen weighting values for the final images. The shaded region indicates the robust values at which the number counts are affected by source confusion.}
    \label{fig:nsource-rob}
\end{figure}
\begin{figure}
    \centering
    \includegraphics[width=0.95\columnwidth]{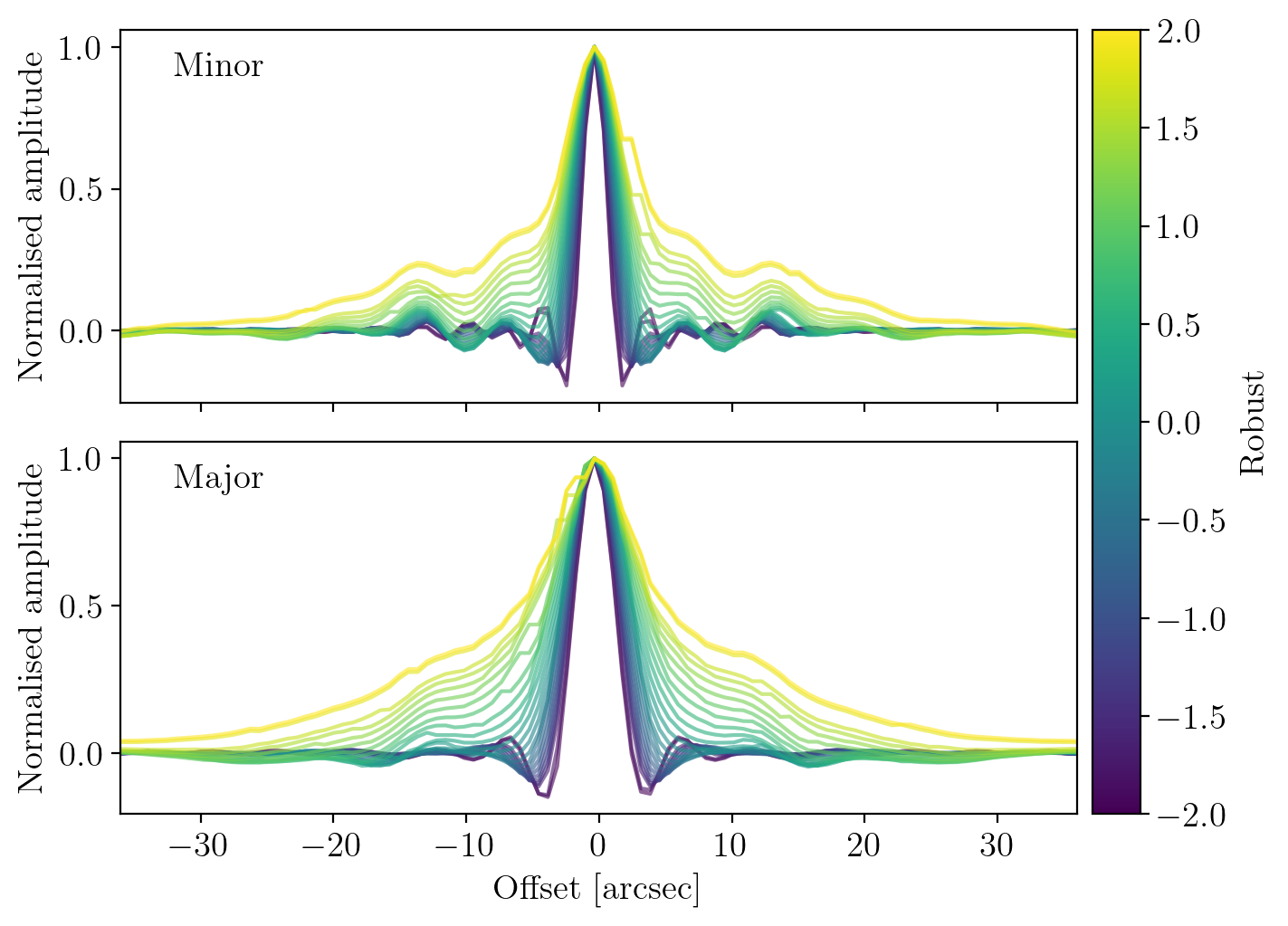}
    \caption{Cross section of the minor (top) and major (bottom) axes of the point spread function, colourised by Briggs robust value.}
    \label{fig:beamsize}
\end{figure}

\section{Differential source counts}\label{sec:sourcecountsapp}
In Section~\ref{sec:sourcecounts}, we present the differential source counts for both robust weightings $R=0.3$ and $R=-0.5$. Table~\ref{tab:sourcecounts} shown the number of sources for each flux density bin, as well as the Euclidean normalised counts and corrected counts for $R = 0.3$ and $R=-0.5$.

\begin{table*}
    \centering
    \caption{Table of source counts for DEEP2 at S-band, for both the $R=0.3$ and $R=-0.5$ images.}
    \small
    \begin{tabular}{l c c c c c c c}
    \hline
    $S$ & $S_{mean}$ & $N$ & Counts & Corrected counts & $N$ & Counts & Corrected counts \\
     & & $R=0.3$ & $R=0.3$ & $R=0.3$ & $R=-0.5$ & $R=-0.5$ & $R=-0.5$\\
    (mJy) & (mJy)   & & (Jy$^{3/2}$ sr$^{-1}$) & (Jy$^{3/2}$ sr$^{-1}$) & & (Jy$^{3/2}$ sr$^{-1}$) & (Jy$^{3/2}$ sr$^{-1}$) \\
    \hline
    0.022 - 0.028 & 0.025 & $31\pm5$ & $0.0264\pm0.0047$ & $0.656\pm0.288$ & $0$ & -- & -- \\ 
    0.028 - 0.036 & 0.032 & $74\pm8$ & $0.0925\pm0.0108$ & $1.28\pm0.42$ & $5\pm2$ & $0.00625\pm0.00280$ & $0.495\pm0.495$\\ 
    0.036 - 0.046 & 0.041 & $127\pm11$ & $0.233\pm0.021$ & $1.98\pm0.53$ & $20\pm4$ & $0.0367\pm0.0082$ & $1.43\pm1.02$\\ 
    0.046 - 0.06 & 0.053 & $141\pm11$ & $0.38\pm0.03$ & $2.25\pm0.48$ & $53\pm7$ & $0.143\pm0.020$ & $2.26\pm1.08$\\ 
    0.06 - 0.077 & 0.069 & $141\pm11$ & $0.557\pm0.047$ & $2.48\pm0.47$ & $66\pm8$ & $0.261\pm0.032$ & $2.51\pm0.82$\\ 
    0.077 - 0.1 & 0.089 & $127\pm11$ & $0.737\pm0.065$ & $2.63\pm0.46$ & $72\pm8$ & $0.418\pm0.049$ & $2.74\pm0.71$\\ 
    0.1 - 0.13 & 0.11 & $101\pm10$ & $0.86\pm0.09$ & $2.57\pm0.42$ & $76\pm8$ & $0.647\pm0.074$ & $3.17\pm0.71$\\ 
    0.13 - 0.17 & 0.15 & $97\pm9$ & $1.21\pm0.123$ & $3.08\pm0.47$ & $59\pm7$ & $0.738\pm0.096$ & $2.89\pm0.60$\\ 
    0.17 - 0.22 & 0.19 & $51\pm7$ & $0.936\pm0.131$ & $2.01\pm0.34$ & $48\pm6$ & $0.881\pm0.127$ & $2.70\pm0.54$\\ 
    0.22 - 0.28 & 0.25 & $54\pm7$ & $1.45\pm0.20$ & $2.76\pm0.44$ & $46\pm6$ & $1.24\pm0.18$ & $3.34\pm0.63$\\ 
    0.28 - 0.36 & 0.32 & $39\pm6$ & $1.54\pm0.25$ & $2.61\pm0.46$ & $35\pm5$ & $1.38\pm0.23$ & $3.23\pm0.63$\\ 
    0.36 - 0.46 & 0.41 & $40\pm6$ & $2.32\pm0.37$ & $3.52\pm0.60$ & $33\pm5$ & $1.92\pm0.33$ & $3.87\pm0.75$\\ 
    0.46 - 0.6 & 0.53 & $27\pm5$ & $2.30\pm0.44$ & $3.20\pm0.65$ & $23\pm4$ & $1.96\pm0.41$ & $3.52\pm0.78$\\ 
    0.6 - 0.77 & 0.69 & $29\pm5$ & $3.63\pm0.67$ & $4.73\pm0.91$ & $25\pm5$ & $3.13\pm0.63$ & $5.12\pm1.12$\\ 
    0.77 - 1 & 0.89 & $20\pm4$ & $3.67\pm0.82$ & $4.51\pm1.03$ & $15\pm3$ & $2.75\pm0.71$ & $4.11\pm1.10$\\ 
    1 - 1.3 & 1.1 & $17\pm4$ & $4.58\pm1.11$ & $5.43\pm1.33$ & $20\pm4$ & $5.39\pm1.20$ & $7.45\pm1.73$\\ 
    1.3 - 1.7 & 1.5 & $22\pm4$ & $8.70\pm1.85$ & $10.0\pm2.2$ & $16\pm4$ & $6.33\pm1.58$ & $8.22\pm2.11$\\ 
    1.7 - 2.2 & 1.9 & $7\pm2$ & $4.06\pm1.54$ & $4.57\pm1.73$ & $10\pm3$ & $5.8\pm1.84$ & $7.19\pm2.30$\\ 
    2.2 - 2.8 & 2.5 & $6\pm2$ & $5.11\pm2.09$ & $5.69\pm2.33$ & $9\pm3$ & $7.67\pm2.56$ & $9.14\pm3.07$\\ 
    2.8 - 3.6 & 3.2 & $6\pm2$ & $7.50\pm3.06$ & $8.23\pm3.37$ & $3\pm1$ & $3.75\pm2.17$ & $4.33\pm2.50$\\ 
    3.6 - 4.6 & 4.1 & $11\pm3$ & $20.2\pm6.1$ & $21.9\pm6.7$ & $11\pm3$ & $20.2\pm6.1$ & $22.9\pm6.9$\\ 
    4.6 - 6 & 5.3 & $15\pm3$ & $40.4\pm10.4$ & $43.6\pm11.3$ & $14\pm3$ & $37.7\pm10.1$ & $42.1\pm11.3$\\ 
    6 - 7.7 & 6.9 & $4\pm2$ & $15.8\pm7.9$ & $16.9\pm8.5$ & $3\pm1$ & $11.9\pm6.9$ & $13.0\pm7.5$\\ 
    7.7 - 10 & 8.9 & $2\pm1$ & $11.6\pm8.2$ & $12.3\pm8.7$ & $2\pm1$ & $11.6\pm8.2$ & $12.3\pm8.7$\\ 
    10 - 13 & 11 & $1\pm1$ & $8.52\pm8.52$ & $8.97\pm8.97$ & $1\pm1$ & $8.52\pm8.52$ & $8.99\pm8.99$\\ 
    13 - 17 & 15 & $2\pm1$ & $25.0\pm17.7$ & $26.3\pm18.6$ & $3\pm1$ & $37.5\pm21.7$ & $39.3\pm22.7$\\ 
    17 - 22 & 19 & $3\pm1$ & $55.1\pm31.8$ & $57.9\pm33.4$ & $1\pm1$ & $18.4\pm18.4$ & $19.3\pm19.3$\\ 
    22 - 28 & 25 & $2\pm1$ & $53.9\pm38.1$ & $56.7\pm40.1$ & $1\pm1$ & $26.9\pm26.9$ & $28.3\pm28.3$\\ 
    \hline
    \end{tabular}
    \label{tab:sourcecounts}
\end{table*}

\section{Catalogue}\label{sec:cat-eg}
In Section~\ref{sec:catalogue}, we present the MeerKAT DEEP2 catalogue at S-band frequencies. In Table~\ref{tab:table_desc}, we describe each column of the catalogue in detail, and in Table~\ref{tab:catalogue_entries}, we provide an example of ten random entries in the catalogue. The full catalogue can be found online on the SARAO archive at \url{https://doi.org/10.48479/zdyz-8342}.

\begin{table*}
    \centering
    \caption{Description of catalogue columns.}
    \begin{tabular}{c l l}
    \hline
    Index & Name & Description \\
    \hline 
    1 &  Source\_name & Name of the source following IAU convention (JHHMMSS.ss$\pm$HHMMSS.s). \\
    2 & RA & Right ascension (J2000) of the source. \\
    3 & E\_RA & Error (1$\sigma$) on right ascension of the source. \\
    4 & DEC & Declination (J2000) of the source. \\
    5 & E\_DEC & Error (1$\sigma$) on declination of the source. \\
    6 & Sep\_PC & Distance of source from the pointing centre. \\
    7 & Total\_flux & Integrated flux density of the source based on Gaussian component fits. \\
    8 & E\_Total\_flux & Error (1$\sigma$) on Total\_flux. \\
    9 & Peak\_flux & Peak flux density of the source. \\
    10 & E\_Peak\_flux & Error (1$\sigma$) on peak flux density of the source. \\
    11 & Maj & FWHM of the major axis of the source. \\
    12 & E\_Maj & Error (1$\sigma$) on FWHM of the major axis of the source. \\
    13 & Min & FWHM of the minor axis of the source. \\
    14 & E\_Min & Error (1$\sigma$) on FWHM of the minor axis of the source. \\
    15 & PA & Position angle of the major axis of the source. \\
    16 & E\_PA & Error (1$\sigma$) on position angle of the major axis of the source. \\
    17 & DC\_Maj & FWHM of the deconvolved major axis of the source. \\
    18 & E\_DC\_Maj & Error (1$\sigma$) on FWHM of the deconvolved major axis of the source. \\
    19 & DC\_Min & FWHM of the deconvolved minor axis of the source. \\
    20 & E\_DC\_Min & Error (1$\sigma$) on FWHM of the deconvolved minor axis of the source. \\
    21 & DC\_PA & Position angle of the deconvolved major axis of the source. \\
    22 & E\_DC\_PA & Error (1$\sigma$) on position angle of the deconvolved major axis of the source. \\
    23 & Isl\_Total\_flux & Integrated flux density of the island on which the source is located. \\
    24 & E\_Isl\_Total\_flux &  Error (1$\sigma$) on Isl\_Total\_flux. \\
    25 & Isl\_rms & Average background rms of the island on which the source is located. \\
    26 & Isl\_mean & Average background mean of the island on which the source is located. \\
    27 & S\_Code & Code defining the source structure. `S' = fit by a single Gaussian component. `M' = fit by multiple \\ & &  Gaussian components. `C' = fit by a single Gaussian component on an island that contains other sources. \\
    28 & Resolved & Boolean indicating whether the source is resolved, as determined in Section~\ref{sec:resolved}. \\
    29 & Matthews\_idx$^a$ & Indices of sources from the \citet{Matthews_2021} catalogue to which this source has been matched. \\
    30 & Matthews\_flux$^a$ & Sum of integrated flux densities of all sources from the \citet{Matthews_2021} catalogue to \\ & &  which this source has been matched. \\
    31 & Alpha\_LS$^b$ & Spectral index between L- and S-band of the source. \\
    32 & E\_Alpha\_LS$^b$ & Error (1$\sigma$) on spectral index between L- and S-band of the source. \\
    33 & Alpha\_S$^b$ & In-band spectral index of the source. \\
    34 & E\_Alpha\_S$^b$ & Error (1$\sigma$) on in-band spectral index of the source. \\
    \hline
    \multicolumn{3}{l}{$^a$ Cross-matching with L-band sources has only been performed for sources in the R=0.3 catalogue.} \\
    \multicolumn{3}{l}{$^b$ Spectral indices have only been determined for sources in the R=0.3 catalogue detected above 10$\sigma$.} \\
    \end{tabular}
    \label{tab:table_desc}
\end{table*}

\begin{table*}
    \centering
    \caption{Catalogue entries for ten random sources out of the DEEP2 R=0.3 S-band catalogue. The full catalogues can be obtained at \url{https://doi.org/10.48479/zdyz-8342}.}
    \begin{tabular}{c c c c c c c c c c c c}
    \hline
    Source\_name & RA & E\_RA & DEC & E\_DEC & Sep\_PC & Total\_flux & E\_Total\_flux & Peak\_flux & E\_Peak\_flux \\
     & (deg) & (deg) & (deg) & (deg) & (deg) & (mJy) & (mJy) & (mJy/beam) & (mJy/beam) \\
    (1) & (2) & (3) & (4) & (5) & (6) & (7) & (8) &  (9) & (10) \\
    \hline 
    J040802.06-794945.6 & 62.008568 & 0.000157 & -79.829332 & 0.000439 & 0.292 & 0.049 & 0.023 & 0.027 & 0.008 \\
    J041121.40-794623.8 & 62.839161 & 0.000059 & -79.773264 & 0.000187 & 0.244 & 0.037 & 0.012 & 0.041 & 0.007 \\
    J041211.94-800800.2 & 63.049736 & 0.000201 & -80.133381 & 0.000157 & 0.144 & 0.143 & 0.015 & 0.066 & 0.006 \\
    J041245.23-795936.8 & 63.188479 & 0.000033 & -79.993568 & 0.000101 & 0.030 & 0.056 & 0.009 & 0.054 & 0.005 \\
    J041255.95-802105.5 & 63.233125 & 0.000048 & -80.351540 & 0.000099 & 0.352 & 0.091 & 0.017 & 0.093 & 0.009 \\
    J041323.88-801419.4 & 63.349503 & 0.000045 & -80.238712 & 0.000137 & 0.239 & 0.057 & 0.013 & 0.054 & 0.007 \\
    J041427.55-795357.0 & 63.614784 & 0.000087 & -79.899159 & 0.000130 & 0.110 & 0.027 & 0.009 & 0.029 & 0.005 \\
    J041639.40-802053.5 & 64.164161 & 0.000035 & -80.348191 & 0.000104 & 0.374 & 0.139 & 0.022 & 0.124 & 0.011 \\
    J041813.39-795449.4 & 64.555785 & 0.000102 & -79.913723 & 0.000145 & 0.226 & 0.111 & 0.019 & 0.057 & 0.007 \\
    J042228.16-800416.9 & 65.617353 & 0.000010 & -80.071365 & 0.000026 & 0.397 & 0.534 & 0.023 & 0.528 & 0.013 \\
    \hline
    \end{tabular}
    \begin{tabular}{| c c c c c c c c c c c c c}
    \hline
    Maj & E\_Maj & Min & E\_Min & PA & E\_PA & DC\_Maj & E\_DC\_Maj & DC\_Min & E\_DC\_Min & DC\_PA & E\_DC\_PA \\
    (arcsec) & (arcsec) & (arcsec) & (arcsec) & (deg) & (deg) & (arcsec) & (arcsec) & (arcsec) & (arcsec) & (deg) & (deg) \\
    (11) & (12) & (13) & (14) & (15) & (16) & (17) & (18) & (19)  & (20) & (21) & (22) \\
    \hline
    9.620 & 3.724 & 5.242 & 1.319 & 178.9 & 30.3  & 6.674 & 3.724 & 3.317 & 1.319 & 7.7   & 30.3  \\
    6.949 & 1.616 & 3.532 & 0.404 & 169.7 & 13.0  & 0.000 & 1.616 & 0.000 & 0.404 & 0.0   & 13.0  \\
    9.939 & 2.083 & 4.738 & 0.573 & 37.1  & 12.7  & 8.471 & 2.083 & 0.000 & 0.573 & 48.8  & 12.7  \\
    7.326 & 0.867 & 3.840 & 0.241 & 170.6 & 7.2   & 0.000 & 0.867 & 0.000 & 0.241 & 0.0   & 7.2   \\
    6.746 & 0.881 & 3.955 & 0.301 & 161.2 & 9.5   & 0.000 & 0.881 & 0.000 & 0.301 & 0.0   & 9.5   \\
    7.166 & 1.161 & 4.066 & 0.381 & 178.1 & 11.2  & 2.673 & 1.161 & 0.000 & 0.381 & 38.1  & 11.2  \\
    5.629 & 1.138 & 4.381 & 0.681 & 162.3 & 31.7  & 0.000 & 1.138 & 0.000 & 0.681 & 0.0   & 31.7  \\
    7.532 & 0.892 & 4.051 & 0.267 & 170.1 & 6.7   & 2.727 & 0.892 & 1.114 & 0.267 & 0.7   & 6.7   \\
    9.716 & 1.406 & 5.460 & 0.532 & 147.0 & 10.4  & 7.238 & 1.406 & 2.711 & 0.532 & 131.7 & 10.4  \\
    7.061 & 0.228 & 3.898 & 0.070 & 165.4 & 179.9 & 0.000 & 0.228 & 0.000 & 0.070 & 0.0   & 179.9 \\
    \hline
    \end{tabular}
    \begin{tabular}{c c c c c c c c c}
    \hline
    Isl\_Total\_flux & E\_Isl\_Total\_flux & Isl\_rms & Isl\_mean & S\_Code & Resolved & Matthews\_idx & Matthews\_flux & Alpha\_LS \\
    (mJy) & (mJy) & (mJy/beam) & (mJy/beam) & & & & (mJy) &  \\
    (23) & (24) & (25) & (26) & (27) & (28) & (29) & (30) & (31) \\
    \hline 
    0.037 & 0.008 & 0.008 & 1.216e-4  & S & False & 4062      & 0.101 & -- \\
    0.026 & 0.006 & 0.008 & -5.406e-4 & S & False & 6637      & 0.058 & -- \\
    0.121 & 0.012 & 0.006 & -3.955e-4 & M & True  & 7323,7359 & 0.204 & -- \\
    0.050 & 0.006 & 0.005 & -6.689e-5 & S & False & 7768      & 0.086 & -- \\
    0.075 & 0.011 & 0.010 & 4.017e-4  & S & False & 7915      & 0.219 & -- \\
    0.046 & 0.007 & 0.007 & 4.868e-5  & S & False & 8283      & 0.087 & -- \\
    0.019 & 0.004 & 0.005 & 1.738e-4  & S & False & 9213      & 0.027 & -- \\
    0.117 & 0.013 & 0.011 & 9.564e-5  & S & False & 11113     & 0.225 & -- \\
    0.086 & 0.009 & 0.007 & -2.584e-4 & S & True  & 12408     & 0.141 & -- \\
    0.526 & 0.018 & 0.013 & 1.608e-3  & S & False & 15507     & 0.948 & -0.355 \\
    \hline
    \end{tabular}
    \begin{tabular}{c c c}
    \hline
     E\_Alpha\_LS & Alpha\_S & E\_Alpha\_S \\
     (31) & (33) & (34) \\
    \hline
     -- & -- & -- \\
     -- & -- & -- \\
     -- & -- & -- \\
     -- & -- & -- \\
     -- & -- & -- \\
     -- & -- & -- \\
     -- & -- & -- \\
     -- & -- & -- \\
     -- & -- & --  \\
     0.156 & 0.169 & 0.462 \\
    \hline
    \end{tabular}
    \label{tab:catalogue_entries}
\end{table*}


\bsp	
\label{lastpage}
\end{document}